# Visualizing the Impact of Quenched Disorder on 2D Electron Wigner Solids

Zhehao Ge[1,*,†], Conor Smith[2,3,*], Zehao He[1,4,5,*], Yubo Yang[2,6], Qize Li[1,7], Ha-Leem Kim[1,4,8], Ziyu Xiang[1,4,7,8], Jianghan Xiao[1,4,7], Wenjie Zhou[1], Salman Kahn[1,4,8], Aining Hu[1], Melike Erdi[9], Rounak Banerjee[9], Takashi Taniguchi[10], Kenji Watanabe[11], Seth Ariel Tongay[9], Miguel A. Morales[2], Shiwei Zhang[2], Feng Wang[1,4,8,†], Michael F. Crommie[1,4,8,†]

[1]*Department of Physics, University of California, Berkeley, Berkeley, CA, USA*

[2]*Center for Computational Quantum Physics, Flatiron Institute, New York, NY, USA*

[3]*Department of Electrical and Computer Engineering, University of New Mexico, Albuquerque, NW, USA*

[4]*Materials Sciences Division, Lawrence Berkeley National Laboratory, Berkeley, CA, USA*

[5]*Department of Material Science and Engineering, University of California, Berkeley, Berkeley, CA, USA*

[6]*Department of Physics and Astronomy, Hofstra University, Hempstead, NY, USA*

[7]*Graduate Group in Applied Science and Technology, University of California, Berkeley, Berkeley, CA, USA*

[8]*Kavli Energy Nano Sciences Institute at the University of California Berkeley and the Lawrence Berkeley National Laboratory, Berkeley, CA, USA*

[9]*School for Engineering of Matter, Transport and Energy, Arizona State University, Tempe, AZ, USA*

[10]*Research Center for Materials Nanoarchitectonics, National Institute for Materials Science, Tsukuba, Japan*

[11]*Research Center for Electronic and Optical Materials, National Institute for Materials Science, Tsukuba, Japan*

* These authors contributed equally to this work.

[†]Email: zge2@berkeley.edu, fengwang76@berkeley.edu, crommie@berkeley.edu

# Abstract:

Electron Wigner solids (WSs)[1-12] provide an ideal system for understanding the competing effects of electron-electron and electron-disorder interactions, a central unsolved problem in condensed matter physics. Progress in this topic has been limited by a lack of single-defect-resolved experimental measurements as well as accurate theoretical tools to enable realistic experiment/theory comparison. Here we overcome these limitations by combining atomically-resolved scanning tunneling microscopy (STM) with neural-quantum-state quantum Monte Carlo (NQS-QMC) simulation of disordered 2D electron WSs to discover new disorder-induced physical regimes of correlated electron behavior. STM was used to image the electron density ($n_e$) dependent evolution of electron WSs in gate-tunable bilayer $MoSe_2$ devices with varying long-range ($n_{LR}$) and short-range ($n_{SR}$) disorder densities. These images were compared to NQS-QMC simulations using realistic disorder maps extracted from experiment, thus allowing the roles of different disorder types to be disentangled. We identify two distinct physical regimes for disordered electron WSs that depend on $n_{SR}$. For $n_{SR} \lesssim n_e$ the WS behavior is dominated by long-range disorder and features extensive mixed solid-liquid phases, a new type of local re-entrant melting/crystallization, and prominent Friedel oscillations. In contrast, when $n_{SR} \gg n_e$ these features are suppressed and a more robust amorphous WS phase emerges that persists to higher $n_e$, highlighting the importance of short-range disorder in this regime. Our work establishes a powerful framework for studying disordered quantum solids via a combined experimental-theoretical approach.

Electrons in clean 2D systems transition from a Fermi liquid to a Wigner crystal (WC) when electron-electron interactions dominate over kinetic energy at low electron density ($n_e$)[13-18]. Quantum Monte Carlo (QMC) simulations predict that this transition occurs when the Wigner–Seitz radius ($r_s$, defined as half the average inter-electron distance in units of the effective Bohr radius) exceeds ~37[14,17]. In real materials disorder is inevitable and significantly alters correlated electron behavior[1-12], thus leading to difficulties in quantitatively modeling experimental results. Transport and optics experiments[19-22], for example, suggest that quenched disorder reduces critical $r_s$ values for the solid-liquid phase transition in 2D electronic systems. The physical mechanisms behind these experimental observations are still debated[9-12,23] due to the poorly understood interplay between electron-electron and electron-disorder interactions. Uncertainty here is fueled by an inability to experimentally evaluate precise disorder configurations in transport and optics measurements, a significant impediment to realistic experiment/theory comparison. Scanning tunneling microscopy (STM) provides a chance to overcome this limitation through its ability to evaluate both microscopic electron distributions[24-28] *and* atomic-scale defect configurations. Combined with advances in QMC methods[17] to simulate interacting electrons and accurately treat disorder, this creates new opportunities for advancing our fundamental understanding of how strongly correlated electron systems behave in the presence of defects.

Here we employ STM techniques to directly visualize the impact of quenched disorder on electron Wigner solids (WSs) in gate-tunable bilayer $MoSe_2$ (BL-$MoSe_2$) devices and compare these results to QMC simulations that incorporate realistic defect configurations (the term WS distinguishes Coulomb-driven electron solids in the presence of disorder from idealized Wigner crystals[28-31]). Atomically-resolved STM imaging allows us to map charged and neutral (i.e., isovalent) atomic defects and simultaneously measure local electron density with sub-nanometer spatial resolution. While previous hole-based WSs in BL-$MoSe_2$ could not be fully melted in experiment[28], our electron WSs melt readily due to the smaller conduction-band effective mass[28,32]. We observe two new regimes of 2D electronic behavior that depend on the density of "short-range" disorder ($n_{SR}$) arising from isovalent defects (long-range disorder density ($n_{LR}$) due to *charged* defects in our measurements is generally smaller than $n_e$). The first regime is the "low defect density" (LDD) regime and occurs when $n_{SR} \lesssim n_e$. Here WS behavior is dominated by charged defects and the effects of isovalent defects are negligible. The LDD regime exhibits extensive mixed-state coexistence of electron solid and liquid phases, including a novel form of local re-entrant melting/crystallization, as well as prominent Friedel oscillations at higher $n_e$. The second regime is the "high defect density" (HDD) regime and occurs when $n_{SR} \gg n_e$. Here isovalent defects have a large effect and cause novel, emergent WS behavior. The HDD regime exhibits a robust, amorphous WS phase that persists to much higher $n_e$ values than seen for the LDD regime, and it shows no local re-entrant crystallization behavior. Mixed-state behavior is suppressed in the HDD regime and Friedel oscillations are quenched.

These findings were facilitated by our use of a new approach to performing ab initio QMC simulations that treats both electron interactions and disorder accurately, with disorder configurations directly extracted from STM measurements. We overcome the limitations of conventional QMC which relies on distinct variational wavefunction ansatz for different phases and has difficulty capturing the extensive mixed-phase behavior seen in experiment[22,28]. Instead we employ a recently developed *multiple-plane-waves-message-passing neural quantum states* ($(MP)^2$-NQSs) ansatz[17] that enables accurate representation of both solid and liquid phases in a single, unified framework. This approach allows a highly accurate QMC simulation of the quantum melting process of realistic disordered electron WSs for the first time. Our simulations capture both WS and mixed-phase behaviors, allowing clear identification of new defect-induced LDD and HDD regimes in both theory and experiment.

## Disorder Characterization

As shown in Fig. 1a, our BL-$MoSe_2$ devices consist of graphite nanoribbon contacts placed on top of a BL-$MoSe_2$/hexagonal boron nitride (hBN) heterostructure on top of a $SiO_2$/Si substrate (sample fabrication details in SI section S1). The graphite nanoribbon contacts reduce contact resistance to BL-$MoSe_2$ at our STM measurement temperature (~ 4.8 K) and a doped silicon back gate is used to tune $n_e$[25,26,28]. A representative STM topograph of a LDD BL-$MoSe_2$ device is shown in Fig. 1b, revealing various defects that can be classified as either charged or isovalent based on their distinct appearances in STM images. Figure 1c (a close-up of the boxed area in Fig. 1b) shows that charged defects (red circles) exhibit significantly more spatially extended features than isovalent defects (yellow circles), consistent with previous STM studies of TMD layers grown on conducting substrates[33-35]. These two defect types show markedly different electronic signatures in scanning tunneling spectroscopy (STS). For charged defects (Fig. 1d) the BL-$MoSe_2$ conduction band edge shifts upward in bias voltage ($V_S$) over a large distance from the defect, indicating long-range interaction arising from a negative defect charge center[34,35]. In contrast, isovalent defects (Fig. 1e) produce only short-range potentials due to their charge-neutral character. Our images show that the long-range and short-range disorder densities for this region are $n_{LR} \approx 1.6\times10^{11}$ cm$^{-2}$ and $n_{SR} \approx 3.5\times10^{11}$ cm$^{-2}$. (Further characterization of observed defects can be found in SI sections S2 and S3.)

## Low defect density (LDD) behavior

To visualize the behavior of electron WSs in the LDD regime, we employed our recently developed "in-gap" tunneling technique[25,26,28] which minimizes STM tip-induced perturbations (details in SI section S4). Here $V_S$ is adjusted to match the tip-sample work function difference while keeping the tip Fermi level within the BL-$MoSe_2$ band gap (hence the term "in-gap"). Under these conditions the tunnel current arises from conduction band electrons that tunnel to the tip, providing a minimally invasive probe of electron density modulations associated with 2D electron states in BL-$MoSe_2$.

Figure 2a shows an in-gap tunnel current map acquired in the same region as Fig. 1b after tuning the electron density to $n_e \approx 2.9\times10^{11}$ cm$^{-2}$ (see SI section S5 for details on estimating $n_e$). Using an effective mass $m^* \approx 0.54m_e$ (extracted from band structure calculations[28]) and an average background dielectric constant $\varepsilon \approx 2.58\varepsilon_0$ for the hBN/vacuum interface, the experimental $r_s$ at this $n_e$ is estimated to be $r_s = \frac{m^* e^2}{4\pi\varepsilon\hbar^2\sqrt{\pi n_e}} \approx 41.4$ (this $r_s$ value is a crude estimate that ignores complexities arising from the STM tip and GNR contacts, see further discussion in SI section S6). This value exceeds the predicted critical value $r_s \approx 37$ for Wigner crystallization in pristine 2D systems[14,17] and so a WS phase is expected under these conditions. Indeed, a WS is observed experimentally in Fig. 2a where each large bright feature corresponds to a localized electron occupying space between charged defects that are marked by red dashed circles. The electrons are repelled by each other as well as by the stationary charged defects. Long-range crystallinity is disrupted by the disorder, but the system exhibits local patches of distorted triangular lattice, consistent with a disordered WS phase (mesoscopic-scale crystalline order exists in the LDD regime as indicated by peaks seen in structure factors calculated from Fig. 2 (see SI Fig. S8)).

Figures 2b, c show in-gap tunnel current maps of the LDD device acquired at slightly higher $n_e$ where the estimated $r_s$ is lower but remains above the predicted threshold for pristine 2D Wigner crystallization[14,17]. Additional localized electrons are observed for increased $n_e$, as expected. Some regions (marked by yellow arrows in Figs. 2b, c) exhibit apparent delocalization, signaling local WS melting even though $r_s > 37$. Upon further increasing $n_e$ to $\sim 4.0\times10^{11}$ cm$^{-2}$ (Fig. 2d), however, the electrons restabilize into a more well-ordered solid phase. We term this behavior *local re-entrant melting/crystallization* of a WS. Such behavior shows that quenched disorder can induce *non-monotonic* evolution in WSs and qualitatively alter the nature of quantum melting in 2D electronic systems.

Figures 3a-d show in-gap tunnel current maps for the same LDD region as Fig. 2, but for intermediate $n_e$ corresponding to $r_s$ values between 18.3 and 22.5 which are below the predicted Wigner crystallization threshold for pristine systems ($\varepsilon = 3.73\varepsilon_0$ was used here to estimate $r_s$ (see SI section S6)). The WS phase persists across this density range, demonstrating that disorder stabilizes electron crystallization beyond the predicted threshold[14,17] for pristine 2D systems. Such behavior is consistent with predictions that disorder can lower critical $r_s$ values and increase the melting temperature of a WS[1,9,10], as well as with our own temperature and $n_e$-dependent transport measurements for LDD BL-$MoSe_2$ (see SI section S7). Local re-entrant melting/crystallization also continues in this electron density regime. For example, locally melted regions (identified by smeared electron density) shrink from Fig. 3a to Fig. 3b, expand from Fig. 3b to Fig. 3c, and then shrink again from Fig. 3c to Fig. 3d. Melted regions are also observed to coalesce into quasi-one-dimensional (1D) filamentary channels that follow the contours of the long-range disorder potential, showing how the disorder energy landscape influences WS quantum melting dynamics.

Figure 4 shows the high-$n_e$ regime of the LDD sample where WS melting is more pronounced. Here the electron density exhibits a "wavy" liquid pattern between charged defects that is reminiscent of Friedel oscillations[36,37]. In contrast, regions near closely spaced charged defects continue to show localized WS behavior (this *mixed-state* behavior is most apparent in Figs. 4a,b). The transition from a Wigner solid (e.g., Fig. 3d) to a fully melted electron liquid (e.g., Fig. 4d) is accompanied by the emergence of pronounced Friedel oscillations around individual charged defects. Such Friedel oscillations are better seen by plotting the radial average of the local electron density around a representative defect (boxed region in Fig. 4c) at different $n_e$ values (see SI Section S8). A representative experimental density profile is shown in Fig. 4i (red curve) and exhibits decaying oscillations away from the defect. This line-cut differs significantly from the theoretical local density expected for a *noninteracting* Friedel oscillation (green dashed line) that includes no electron correlation effects (see details in SI sections S9 and S10).

## High defect density (HDD) behavior

We now turn to the behavior of disordered WSs in the HDD regime ($n_{SR} \gg n_e$) where isovalent defects play a strong role and the 2D electronic properties differ markedly from the LDD regime. The effects of isovalent defects on WSs are more subtle compared to charged defects because they lack a long-range *1/r* potential but still generate a short-range disorder potential due to chemical and size differences between isovalent impurities and host atoms[38,39]. To investigate the effects of short-range disorder on 2D electron behavior we fabricated a new set of BL-$MoSe_2$ devices using $MoSe_2$ crystals with significantly higher isovalent defect densities than crystals used for LDD devices such as the one shown in Figs. 1-4. The HDD devices (otherwise fabricated identically to LDD devices) exhibit an average of ~28 times higher $n_{SR}$ compared to the LDD devices, while maintaining a comparable $n_{LR}$ (Further analysis of disorder distribution is shown in SI section S3).

Figure 5a shows a representative STM topograph of an HDD device. Only four charged defects are seen in the image (dashed red circles), but numerous isovalent defects can be observed as small black and white specks. Figures 5b-e show in-gap tunneling maps of this same region for $n_e$ values close to those examined for the LDD device in Figs. 2-4. A WS phase is observed for low (Fig. 5b) and intermediate (Fig. 5c) $n_e$ values in the HDD device, similar to the LDD device but with four important differences. First, the local re-entrant melting/crystallization observed in the LDD regime is strongly suppressed in the HDD regime. Second, the WS phase is more robust for HDD devices, persisting to $n_e$ well above values where melting occurs in LDD devices (e.g., $n_e \approx 1.16\times10^{12}$ cm$^{-2}$). Third, the structure factor peaks observed in the LDD regime (Figs. S8) are absent in the HDD regime (SI section S11 and discussion of local bond orientational order in SI section S12). And fourth, the pronounced Friedel oscillations seen in the LDD regime at high $n_e$ are absent in HDD devices. The local density pattern of Fig. 5e remains mostly unchanged at higher $n_e$ values and corresponds more to defect locations than long-range quantum interference or electron correlation effects (see SI section S13).

## NQS-QMC simulation in the LDD regime

We are able to gain insight into the contrasting behaviors observed in the LDD and HDD regimes through neural-network-based quantum Monte Carlo simulations. Such calculations are uniquely well-suited to capture the strong electron correlation effects present in our devices and provide quantitatively precise information for comparison with experiment. Our devices are particularly challenging for computational treatment since they are strongly interacting quantum systems with long-range interactions *and* complex disorder. Even standard QMC approaches that have successfully modeled 2D electrons in semiconductor devices[16,22,40] face severe difficulties accurately capturing disorder effects due to the complicated response of many-body wavefunctions to impurities. The recently developed (MP)2-NQS approach[17] provides us with the necessary accuracy and flexibility to model the electronic behavior of 2D devices in the presence of arbitrary disorder configurations. NQS-based ansatze have recently undergone rapid development to optimize complex many-body wavefunctions as illustrated by recent applications to anomalous Hall crystals[41], the BCS-BEC crossover[42], chiral superconductivity[43], superfluid helium[44], fractional quantum Hall states[45-47], and molecular Wigner crystals[48] among others. By using the NQS-QMC technique together with disorder configurations directly extracted from STM measurements (as described in SI section S14), we are able to accurately model the electronic behavior of our 2D devices and perform quantitative one-to-one comparison between simulation and experiment. This allows robust differentiation between long-range and short-range disorder effects as well as emergent LDD and HDD regimes.

Our simulations for LDD devices at low electron density ($n_e \leq 4\times10^{11}$ cm$^{-2}$) are shown in Figs. 2e-h. The simulations show electrons forming highly localized wavepackets that are repelled by neighboring electrons as well as nearby charged defects, mirroring the experimental behavior shown above (LDD simulations in Figs. 2-4 include only long-range disorder since the influence of short-range disorder in the LDD regime is negligible as shown in SI section S15). At the lowest $n_e$ (Fig. 2e) $r_s$ is well above the threshold for pristine WC formation, confirming our observation of a WS in the presence of disorder. At higher $n_e$ the simulations also show local melting (Fig. 2f, yellow arrow) *and* local re-entrant crystallization (Fig. 2g), similar to experiment. Previous NQS-QMC simulations for a disorder-free 2D electron system[17], in contrast, do not show local re-entrant behavior. Closer examination of experimental and simulated local re-entrant melting/crystallization behavior suggests that this arises from $n_e$-dependent commensuration/incommensuration transitions between electron and disorder distributions, as shown by locking/unlocking between electron density and defect structure factor peaks (details in SI section S16).

Figures 3e-h show QMC simulated electron density maps in the LDD regime at intermediate electron densities ($4\times10^{11}$ cm$^{-2}$ < $n_e$ < $10^{12}$ cm$^{-2}$). Here the agreement between simulation and experiment remains strong, with clear signatures of local re-entrant melting/crystallization in the simulated density maps. Notably, the simulations reveal quasi-1D melted chains that follow the contours of the long-range disorder potential and are strikingly similar to the experimental images above.

Simulating LDD devices at high electron densities ($n_e > 10^{12}$ cm$^{-2}$) is more difficult due to the increased number of electrons, and so we simulate smaller regions. Figures 4e-h show QMC simulations corresponding to the boxed region in Fig. 4a for $n_e$ values corresponding to the experimental data in Figs. 4a-d. The simulations reproduce key experimental features, including mixed-state behavior where highly localized WS electrons near defect clusters coexist with fully melted regions further away. Friedel oscillation formation is also well captured, as demonstrated by the direct theory-experiment comparison in Figs. 4i and S12 for charge density around an isolated charged impurity. The QMC treatment matches the experimental Friedel oscillation better than the noninteracting theoretical expression (green dashed line (see SI section S10)) and is comparable to the calculated electron pair correlation function (black dashed line). The agreement between the experimental Friedel oscillation and QMC simulation emphasizes the importance of electron correlation effects for Friedel oscillations in interacting electronic systems.

## NQS-QMC simulation in the HDD regime

To understand the different role that short-range disorder plays in HDD devices compared to LDD devices, we modeled isovalent defects as screened charges having only a short-range potential. We then systematically examined their effect on QMC simulations as a function of short-range disorder density (while including all charged defects). The positions of the isovalent defects are known from our STM measurements, and so we were able to methodically populate our HDD simulations with an increasing fraction of isovalent defect sites (see details in SI section S14).

Figure 5f shows a simulation of the device in Fig. 5a at $n_e \approx 1.21\times10^{12}$ cm$^{-2}$, but with *no* short-range disorder. Here the local electron density clearly resembles an LDD device at this value of $n_e$ (e.g., Fig. 4a) but looks nothing like the actual HDD device (Fig. 5d). When 1/4 of the effective total number of isovalent defects are added, however, significant differences begin to emerge (Fig. 5g). Some mixed state behavior is still apparent, but electron localization is dramatically enhanced (see SI section S17 for discussion of the importance of electron-electron interactions in this regime). When 1/2 of the effective isovalent defect sites are populated (Fig. 5h) the WS phase becomes more fully developed with minimal signs of local melting and no evidence of Friedel oscillations. Finally, when all effective isovalent defects are included (Fig. 5i) the system exhibits a robust HDD WS phase characterized by uniformly strong electron localization and complete quenching of Friedel oscillations (Friedel oscillations remain quenched up to higher $n_e$ in our simulations, see SI section S18). The resulting electron distribution is strikingly similar to the experimental data shown in Fig. 5d where electrons are well-separated due to Coulomb repulsion and arranged in an amorphous configuration. These results show how a high density of short-range disorder qualitatively alters the nature of the WS phase, both stabilizing it and suppressing crystalline ordering, in contrast to the less stable and more locally ordered WS observed in the LDD regime.

## Conclusion

We have established a powerful experimental-theoretical framework that has enabled the discovery of distinct defect-induced physical regimes for 2D electronic systems exhibiting both strong electron-electron and electron-disorder interactions. In the LDD regime electronic behavior is governed by charged defects and is characterized by local re-entrant WS melting/crystallization, pronounced Friedel oscillations, and negligible influence from isovalent defects. In contrast, the HDD regime exhibits a robust, amorphous WS phase in which *both* local re-entrant melting/crystallization and Friedel oscillations are suppressed. These findings underscore the different roles played by long-range and short-range disorder in interacting 2D systems and show how varying the concentration of short-range disorder can drive new, emergent electronic behavior. While our emphasis is on how such behavior alters the local charge density observed by STM, we expect that it should also affect physical properties measured by other experimental probes. The advances described here should facilitate a deeper understanding of the synergies and competitions that exist between pinning, glassy dynamics, Anderson localization, and strong interactions for correlated materials in the presence of disorder[3,4,9,10,12,49].

## Methods

**Sample fabrication.** The GNR/BL-$MoSe_2$/hBN heterostructures were assembled using a newly developed polymer-based transfer method (see details in SI Section S1). The BL-$MoSe_2$ flakes in LDD and HDD devices were mechanically exfoliated from commercial $MoSe_2$ crystals from HQ Graphene and self-grown $MoSe_2$ crystals, respectively. The Cr/Au contacts were deposited using a shadow-mask with e-beam evaporation. Before STM measurements the completed devices were annealed in UHV at ~340 °C for ~24 hours.

**STM measurements.** The STM measurements were conducted in UHV at pressures less than $1\times10^{-10}$ mbar at 4.8 K in a Createc LT-STM. Electrochemically etched tungsten tips calibrated on Au(111) surface were used in the experiments.

**QMC simulations.** We used the unified $(MP)^2$-NQSs ansatz[17] to describe both Wigner solid and electron liquid states without imposing bias in the trial wavefunction. The ansatz was modified to capture electron disorder interactions by adding an electron-defect Jastrow factor to maintain the cusp condition. The final ansatz included spin-resolved electron-electron Jastrow factors in addition to the all-electron neural-network Jastrow and backflow transformation. The electron coordinates, transformed by the neural-network backflow, were fed into orbitals made up of linear combination of planewaves (see details in SI section S14).

**Acknowledgments:** The authors acknowledge helpful discussions with Ilya Esterlis, Steven Kivelson, Vladimir Calvera, Brian Skinner, Sandeep Joy and Sankar Das Sarma. This work was funded by the US Department of Energy, Office of Science, Basic Energy Sciences, Materials Sciences and Engineering Division under contract DE-AC02-05-CH11231 within the van der Waals heterostructure program KCWF16 (device fabrication, STM measurement). Support was also provided by the Department of Defense Vannevar Bush Faculty Fellowship N00014-23-1-2869 (surface preparation); National Science Foundation award DMR-2221750 (device characterization); and the Flatiron Institute which is a division of the Simons foundation (QMC simulation). Y.Y. acknowledges support from NSF DMR-2532734 (QMC simulation). S.T. acknowledges support from US Department of Energy SC0020653 (excitonic metrology on TMDs crystals), NSF CBET 2330110 (environmental test) and Applied Materials Inc. for defect / dopant analysis. K.W. and T.T. acknowledge support from the JSPS KAKENHI (grants 21H05233 and 23H02052) and World Premier International Research Center Initiative (WPI), MEXT, Japan for hBN crystal fabrication/characterization.

**Author contributions:** Z.G., M.F.C., and F.W. conceived the work and designed the research strategy. Z.H., Q.L., Z.G., and W.Z. fabricated the BL-$MoSe_2$ devices with assistance from S.K. and A.H.. Z.G. and Z.H. carried out STM measurements. C.S., Y.Y., M.A.M. and S.Z. performed NQS-QMC simulations. H.K. performed BL-$MoSe_2$ transport measurements. Z.G., H.Z., Q.L., Z.X., J.X., F.W. and M.F.C. discussed the experiment design and analyzed the experimental data. M.E., R.B., and S.A.T. grew the MoSe2 crystals. K.W. and T.T. grew the hBN crystals. Z.G., M.F.C., F.W., C.S., Y.Y. and S.Z. wrote the paper. All authors commented on the paper.

**Data availability:** Source data are provided with this paper for the main figures. Any additional data that support the findings of this study are available from the corresponding authors upon request.

**Code availability:** All the codes used in this article are available from the corresponding authors upon request.

**Competing interests:** The authors declare no competing interests.

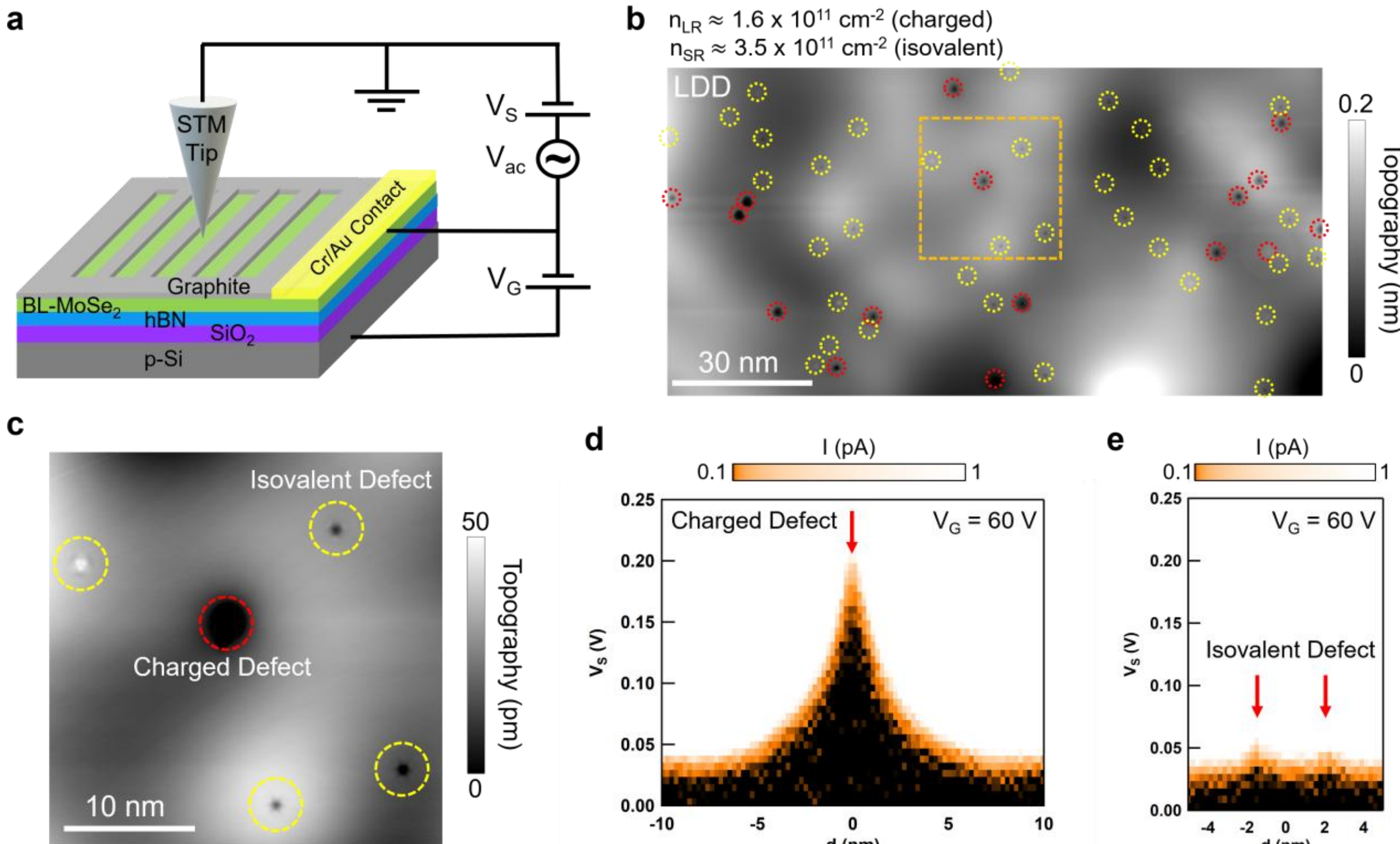

**Figure 1: Experiment setup. a,** Sketch of STM setup for gate-tunable BL-$MoSe_2$ devices. A graphite nanoribbon contact/BL-$MoSe_2$/hBN heterostructure is placed on a 285 nm $SiO_2$/Si substrate. A d.c. bias voltage $V_S$ (with added a.c. modulation $V_{ac}$ for STS) is applied between the STM tip and BL-$MoSe_2$ sample. A backgate voltage $V_G$ is applied between the p-type silicon substrate and the BL-$MoSe_2$. **b,** STM topograph of LDD BL-$MoSe_2$ device. Charged (red circle) and isovalent (yellow circle) defects are marked. Setpoint: $I = 2$ nA, $V_S = 1.09$ V, $V_G = 20$ V. **c,** Close-up topography of the region marked by orange dashed box in (b). **d-e,** $I(V_S, d)$ in log scale color plot measured across (d) a single charged defect and (e) two isovalent defects. Setpoint: $I = 0.1$ nA, $V_S = -1.82$ V, $V_G = 60$ V. The red arrows in (d) and (e) mark defect locations. All STM measurements were performed at $T = 4.8$ K.

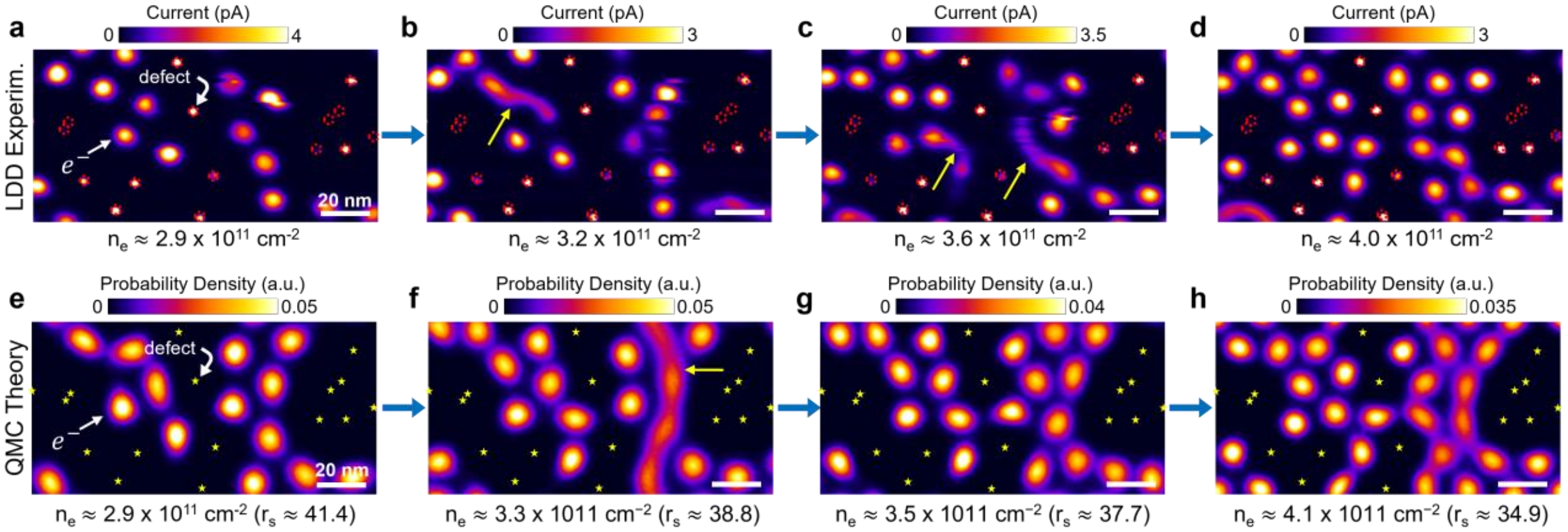

**Figure 2: Wigner solid behavior in LDD regime (low $n_e$). a-d,** Experimental in-gap tunnel current maps of an LDD BL-$MoSe_2$ device measured at (a) $V_G = 1.60$ V, (b) $V_G = 2.05$ V, (c) $V_G = 2.70$ V, and (d) $V_G = 3.30$ V. Setpoint: $V_S = -1.00$ V. Experimental $n_e$ values are shown. Red

dashed circles mark the locations of charged defects. Yellow arrows mark regions exhibiting local electron melting. **e-h,** Theoretical QMC simulations of electron probability distribution for (e) $n_e \approx 2.9\times10^{11}$ cm$^{-2}$, (f) $n_e \approx 3.3\times10^{11}$ cm$^{-2}$, (g) $n_e \approx 3.5\times10^{11}$ cm$^{-2}$, and (h) $n_e \approx 4.1\times10^{11}$ cm$^{-2}$. Yellow stars mark the locations of charged defects obtained from experimental positions and used in the simulations. Yellow arrows mark regions exhibiting local electron melting. Dielectric constant used in QMC simulations for (e)-(h) is $\varepsilon = 2.58\varepsilon_0$.

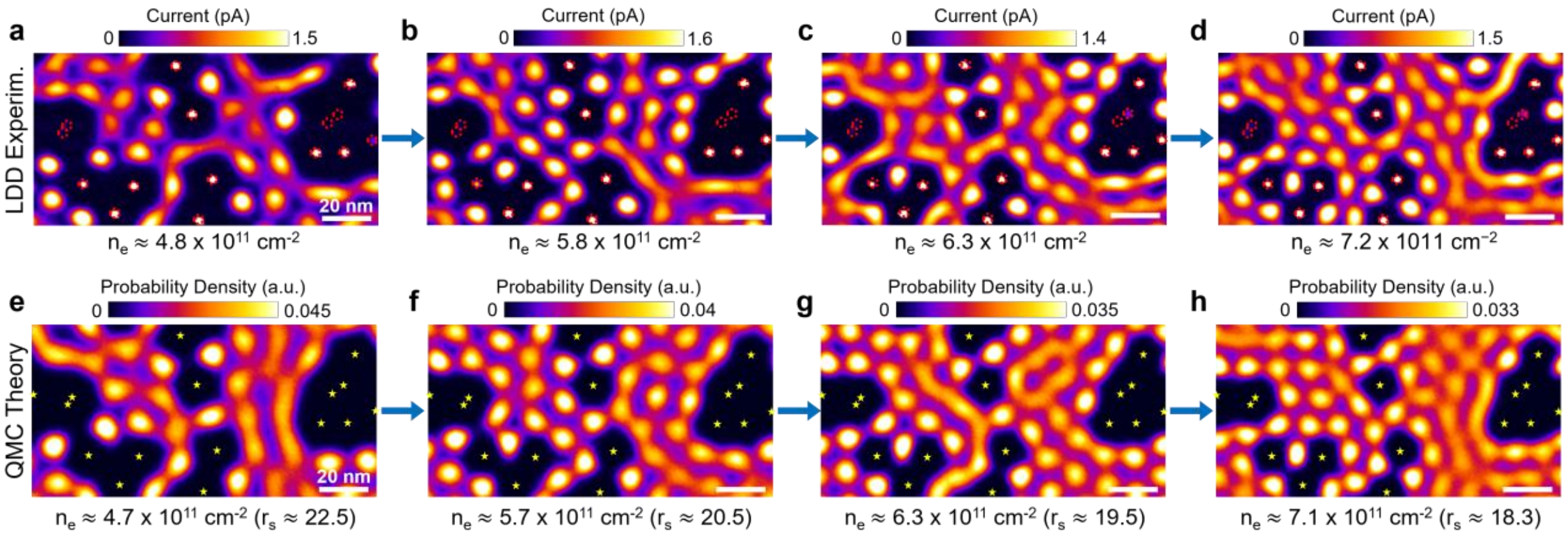

**Figure 3: Wigner solid behavior in LDD regime (intermediate $n_e$). a-d,** Experimental in-gap tunnel current maps of an LDD BL-$MoSe_2$ device measured at (a) $V_G = 4.50$ V, (b) $V_G = 5.80$ V, (c) $V_G = 7.00$ V, and (d) $V_G = 8.15$ V. Setpoint: $V_S = -1.00$ V. Experimental $n_e$ values are shown. Red dashed circles mark the locations of charged defects. **e-h,** Theoretical QMC simulations of electron probability distribution for (e) $n_e \approx 4.7\times10^{11}$ cm$^{-2}$, (f) $n_e \approx 5.7\times10^{11}$ cm$^{-2}$, (g) $n_e \approx 6.3\times10^{11}$ cm$^{-2}$, and (h) $n_e \approx 7.1\times10^{11}$ cm$^{-2}$. Yellow stars mark the locations of charged defects obtained from experimental positions and used in the simulations. Dielectric constant used in QMC simulations for (e)-(h) is $\varepsilon = 3.73\varepsilon_0$.

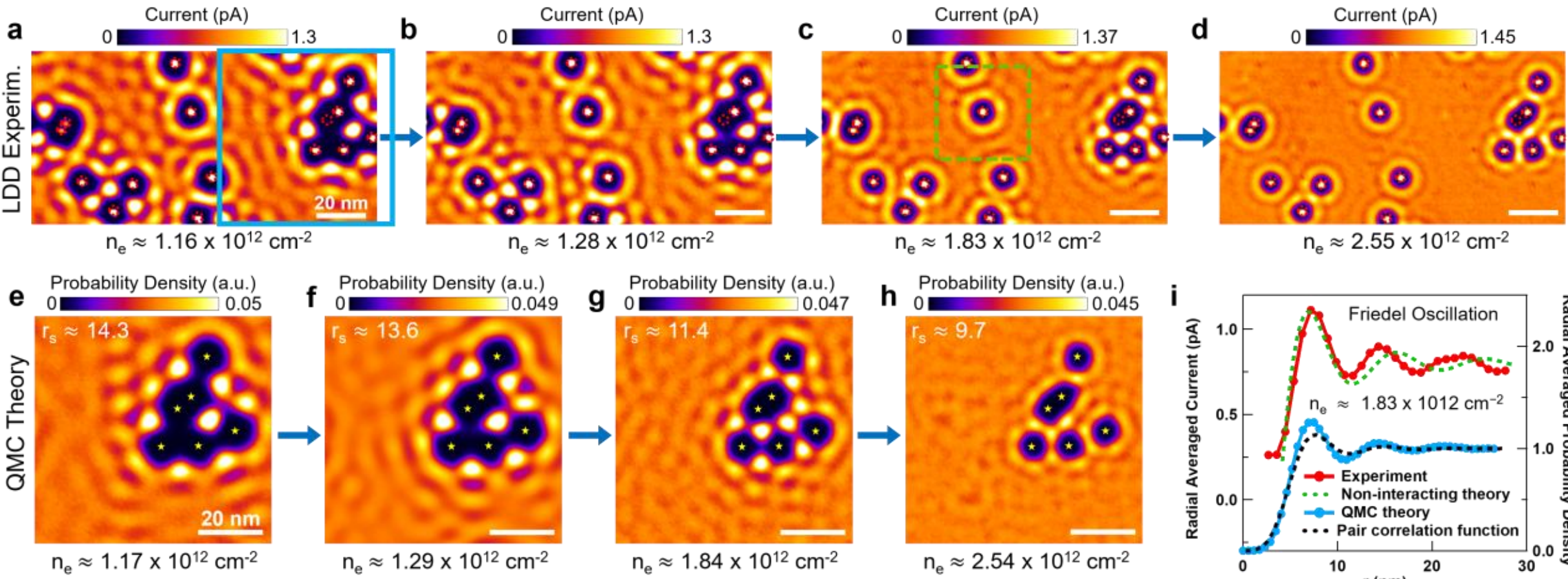

**Figure 4: Electron solid-liquid transition in LDD regime (high $n_e$). a-d,** Experimental in-gap tunnel current maps of an LDD BL-$MoSe_2$ device measured at (a) $V_G = 13.80$ V, (b) $V_G = 15.50$ V, (c) $V_G = 23.00$ V, and (d) $V_G = 33.00$ V. Setpoint: $V_S = -1.00$ V. Experimental $n_e$ values are shown. Red dashed circles mark the locations of charged defects. **e-h,** Theoretical QMC simulations of electron probability distribution for the blue boxed region shown in (a) for (e) $n_e \approx 1.17\times10^{12}$ cm$^{-2}$, (f) $n_e \approx 1.29\times10^{12}$ cm$^{-2}$, (g) $n_e \approx 1.84\times10^{12}$ cm$^{-2}$, and (h) $n_e \approx 2.54\times10^{12}$ cm$^{-2}$.

Yellow stars mark the locations of charged defects obtained from experimental positions and used in the simulations. Dielectric constant used in QMC simulations for (e)-(h) is $\varepsilon = 3.73\varepsilon_0$. **i,** Radial averaged profile of experimental electron density around boxed charged defect in (c) (red curve). Theoretical electron density using non-interacting Friedel oscillation expression (green dashed curve -- details in SI section S10). Bottom curve shows QMC simulation of electron probability density around charged defect at same $n_e$ (blue). Electron pair correlation function calculated using QMC at same $n_e$ (black dashed line). Dielectric constant used in QMC simulations is $\varepsilon = 3.73\varepsilon_0$.

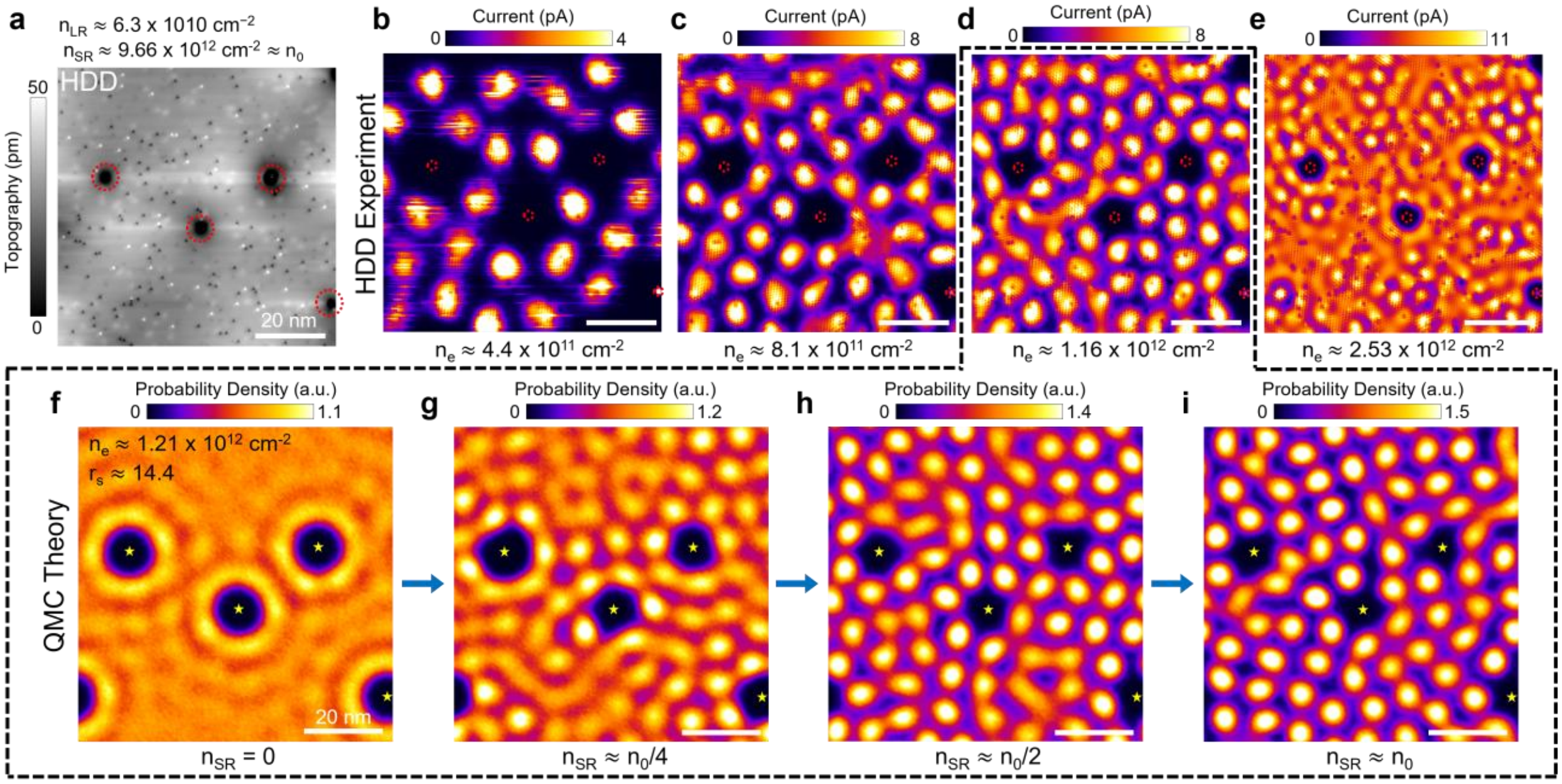

**Figure 5: 2D electronic behavior in HDD regime. a,** STM topograph of HDD BL-$MoSe_2$ device. Red circles mark charged defects. Setpoint: $I = 0.8$ nA, $V_S = 1.09$ V, $V_G = 15$ V. **b-e,** Experimental in-gap tunnel current maps measured at (b) $V_G = 10.00$ V, (c) $V_G = 15.00$ V, (d) $V_G = 20.00$ V, and (e) $V_G = 40.00$ V. Setpoint: $V_S = -1.14$ V. Experimental $n_e$ values are shown. Red dashed circles mark the locations of charged defects. **f-i,** Theoretical QMC simulations of electron probability distribution for (f) $n_{SR} = 0$, (g) $n_{SR} \approx n_0/4$, (h) $n_{SR} \approx n_0/2$, and (i) $n_{SR} \approx n_0$. $n_0 \approx 7.63\times10^{12}$ $cm^{-2}$, $n_e \approx 1.21\times10^{12}$ $cm^{-2}$, $r_s \approx 14.4$. Dielectric constant used in QMC simulations is $\varepsilon = 3.73\varepsilon_0$.

# Supplementary Information for

# Visualizing the Impact of Quenched Disorder on 2D Electron Wigner Solids

Zhehao Ge[1,*,†], Conor Smith[2,3,*], Zehao He[1,4,5,*], Yubo Yang[2,6], Qize Li[1,7], Haleem Kim[1,4,8], Ziyu Xiang[1,4,7,8], Jianghan Xiao[1,4,7], Wenjie Zhou[1], Salman Kahn[1,4,8], Aining Hu[1], Melike Erdi[9], Rounak Banerjee[9], Takashi Taniguchi[10], Kenji Watanabe[11], Seth Ariel Tongay[9], Miguel A. Morales[2], Shiwei Zhang[2], Feng Wang[1,4,8,†], Michael F. Crommie[1,4,8,†]

[1] *Department of Physics, University of California, Berkeley, Berkeley, CA, USA*

[2] *Center for Computational Quantum Physics, Flatiron Institute, New York, NY, USA*

[3] *Department of Electrical and Computer Engineering, University of New Mexico, Albuquerque, NW, USA*

[4] *Materials Sciences Division, Lawrence Berkeley National Laboratory, Berkeley, CA, USA*

[5] *Department of Material Science and Engineering, University of California, Berkeley, Berkeley, CA, USA*

[6] *Department of Physics and Astronomy, Hofstra University, Hempstead, NY, USA*

[7] *Graduate Group in Applied Science and Technology, University of California, Berkeley, Berkeley, CA, USA*

[8] *Kavli Energy Nano Sciences Institute at the University of California Berkeley and the Lawrence Berkeley National Laboratory, Berkeley, CA, USA*

[9] *School for Engineering of Matter, Transport and Energy, Arizona State University, Tempe, AZ, USA*

[10] *Research Center for Materials Nanoarchitectonics, National Institute for Materials Science, Tsukuba, Japan*

[11] *Research Center for Electronic and Optical Materials, National Institute for Materials Science, Tsukuba, Japan*

* These authors contributed equally to this work.

[†] Email: zge2@berkeley.edu, fengwang76@berkeley.edu, crommie@berkeley.edu

## S1. Sample fabrication method

To enable ultra-clean bilayer $MoSe_2$ (BL-$MoSe_2$) surfaces for STM measurements, we developed a "BN-sliding" technique as sketched in Fig. S1 to fabricate BL-$MoSe_2$ devices for our experiments. With this technique the BL-$MoSe_2$ surface does not touch any polymer during the entire sample fabrication process, thus yielding clean sample surface suitable for STM studies. We first use a PVC/PDMS stamp[1] to pick-up the top hBN, the graphite nanoribbon (GNR) contacts, the BL-$MoSe_2$, and the bottom hBN at ~80 °C (Fig. S1a). Then we drop the heterostructure onto a 285nm $SiO_2$/Si substrate at ~130 °C (Fig. S1b). Next, we deposit a 5 nm/60 nm Cr/Au contact onto the GNR using e-beam evaporation (Fig. S1c). Afterwards we use a PVC/PDMS stamp to only touch the top hBN and pull it away from the sample at ~80 °C (Fig. S1d). This leaves the GNR/BL-$MoSe_2$/hBN device with exposed BL-$MoSe_2$ surface on the $SiO_2$/Si substrate (Fig. S1e). Next we perform AFM tip cleaning[2,3] on the exposed BL-$MoSe_2$ regions before transferring the device into UHV. Finally, we anneal the BL-$MoSe_2$ sample in UHV at ~340 °C for ~24 hours before transferring the sample into the STM for measurements.

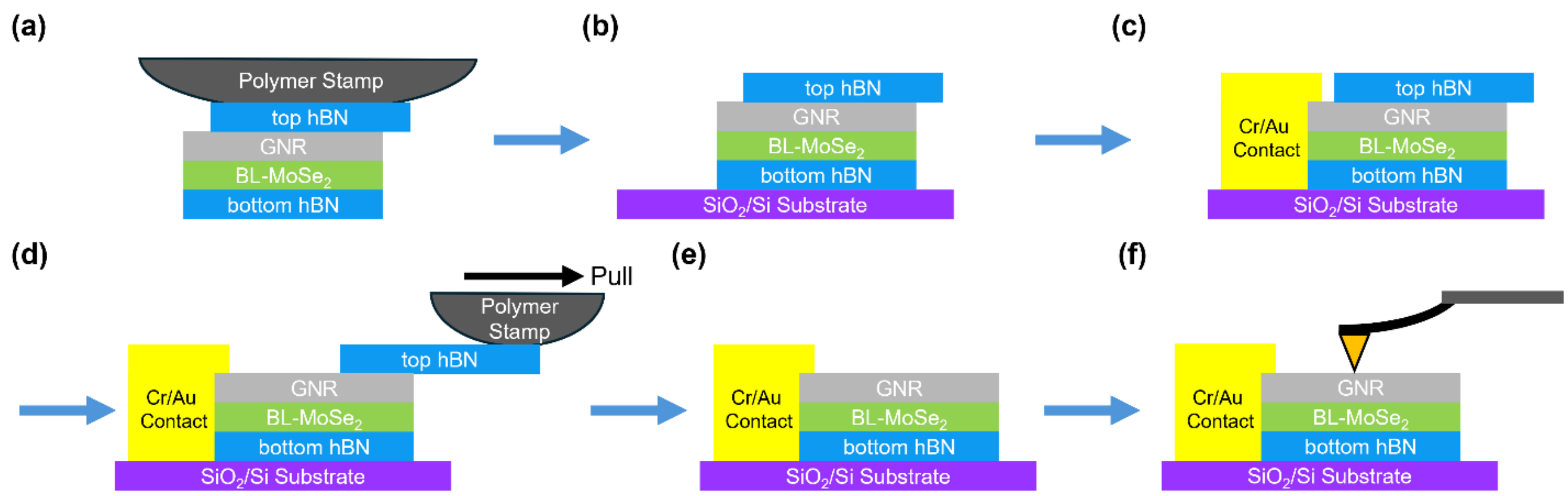

**Figure S1: Schematic of the sample fabrication process for BL-$MoSe_2$ devices used in our experiments.**

**S2. BL-MoSe2 defect characterization**

We acquired high resolution STM topography data of the observed charged and isovalent defects which allowed us to tentatively identify their probable origins. Figure S2a shows high resolution STM topography taken in an HDD region where many different types of $MoSe_2$ defects are visible (marked by yellow dashed circles and labeled A-F). In Fig. S2b we show close-up images of the $MoSe_2$ defects marked in Fig. S2a. By comparing their appearance with STM topography of $WS_2$ defects from prior combined STM/AFM studies[4] we tentatively assign defect A to be a charged defect (CD), defects B and C to be substitutional defects at selenium sites in the bottom/top selenium layer of $MoSe_2$ with Se substituted by oxygen ($O_{Se}$), and defects D and E to be selenium vacancies in the top/bottom selenium layer of $MoSe_2$ ($Vac_{Se}$). Defect F does not look like any prior identified $WS_2$ point defects and so we label it as unidentified. The central conclusions of our work do not rely on the exact atomic structure of these defects. The main point of our work is how the coexistence of long-range charged disorder and short-range neutral disorder together control the pinning, melting, and spatial inhomogeneity of electron Wigner solids.

We have also performed scanning tunneling spectroscopy (STS) on some of the different types of $MoSe_2$ defects, as shown in Fig. S3. By comparing STS taken at defect sites (Figs. S3b-f) with STS at a defect free BL-$MoSe_2$ region (Fig. S3a), we identify a defect state (marked by black arrow) for STS taken on an $O_{Se}$ top defect (Fig. S3c) and charged defect (Fig. S3d). For STS of charge-neutral defects (Figs. S3b,c,e,f) the bias voltage positions of the conduction band edge are more or less consistent with the defect-free region (Fig. S3a). On the charged defect (Fig. S3d), however, the conduction band edge lies at a more positive bias voltage compared to the defect free region (Fig. S3a), due to a band bending effect caused by negative charge localized at the charged defect. To estimate the electron number localized on charged defects we compare the average distance between charged defects and neighboring free electrons (Fig. S4). The (electron)-(charged defect) distance is consistent with the average inter-electron distance extracted through $k_{WS}$. This result shows that the net localized defect electron number on a charged defect is the same as the electron number for a single Wigner solid electron wavepacket (one).

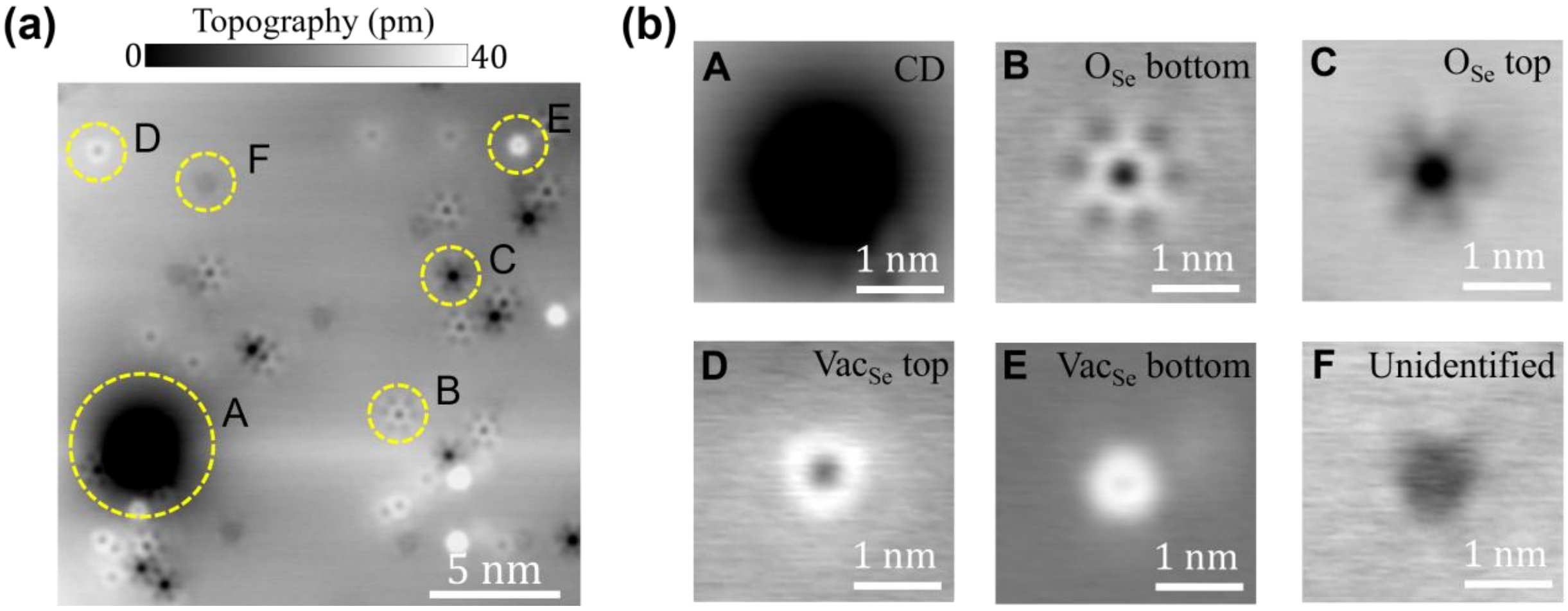

**Figure S2: Identification of possible origins of charged and isovalent defects. a,** STM topography of BL-$MoSe_2$ in the HDD regime. Yellow circles indicate different types of $MoSe_2$ atomic defects observed in this region. The STM topography was taken with a setpoint $I = 0.8$ nA and $V_S = 1.2$ V at $V_G = 15$ V. **b,** Zoomed-in topography of the $MoSe_2$ atomic defects circled in (a).

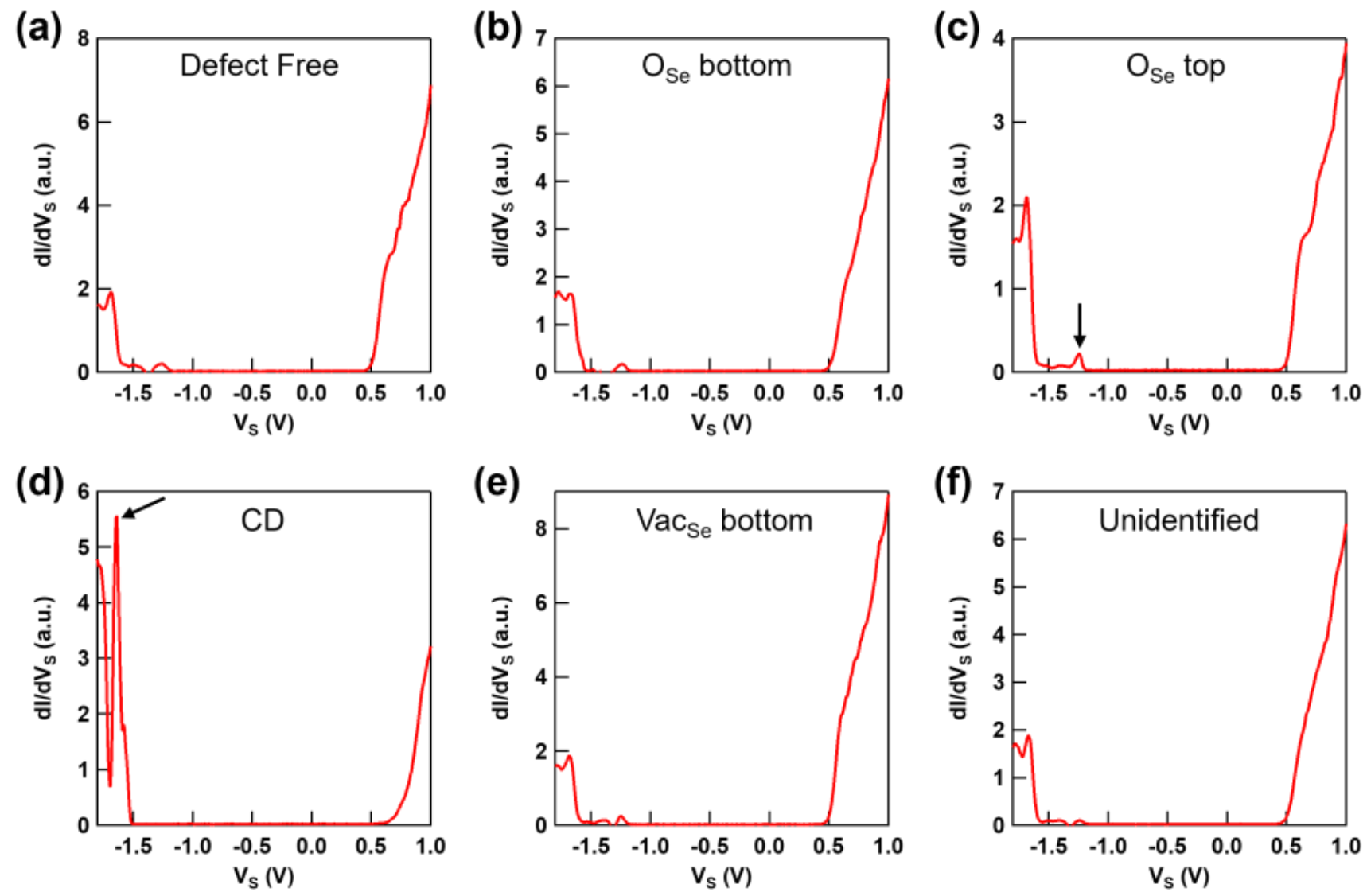

**Figure S3: Scanning tunneling spectroscopy (STS) of different types of BL-MoSe2 defects. a-f,** STS taken at $V_G = 15$ V at (a) a defect-free HDD BL-MoSe2 spot, (b) an $O_{Se}$ bottom defect, (c) a $O_{Se}$ top defect, (d) a charged defect, (e) a $Vac_{Se}$ bottom defect, and (f) an unidentified defect. A set point of $I = 0.15$ nA, $V_S = -1.82$ V with an applied $V_{ac} = 20$ mV was used to acquire the data in (a)-(c) and (e)-(f). A set point of $I = 0.4$ nA, $V_S = 1.09$ V with an applied $V_{ac} = 20$ mV was used to acquire the data in (d).

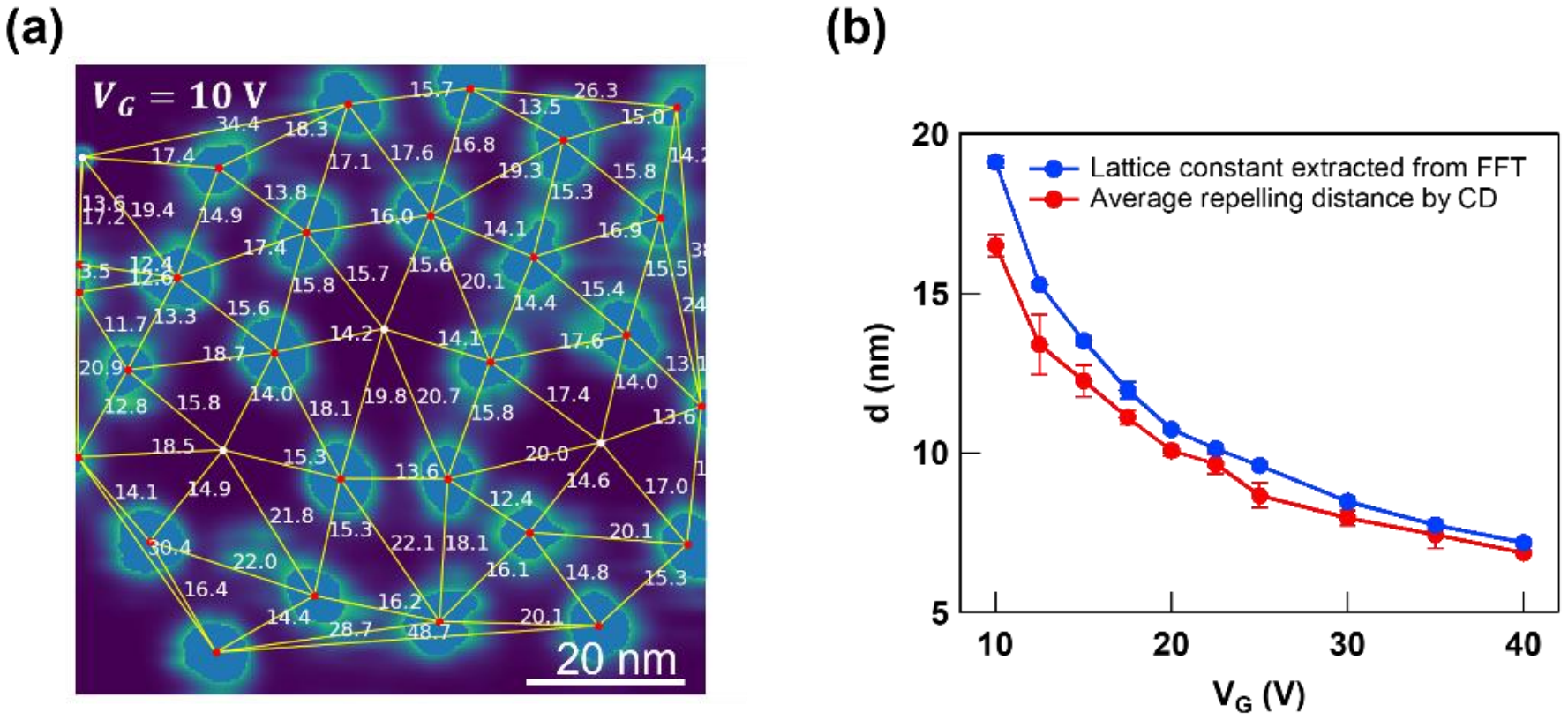

**Figure S4: Gate-dependence of average (electron)-(charged defect) distance. a,** Delaunay triangulation of an electron Wigner solid image in the HDD regime at $V_G = 10$ V. The red dots represent free electron centers. The white dots represent charged defect centers. The white lines and corresponding numbers represent neighboring electron-electron or electron-charged defect distances in units of nm. **b,** Gate-dependence of average neighboring electron-charged defect distance extracted from Delaunay triangulation plot (red line) and inter-electron distance converted from $k_{WS}$ (blue line).

### S3. Disorder distribution for LDD and HDD regions

In Figs. S5a,b we show the defect distributions (including both charged and isovalent defects) extracted from STM topography for the HDD and LDD regions shown in the main text. Their FFT amplitude maps (Figs. S5c,d) lack any clear features, indicating that the defect distributions are random in both the LDD and HDD regimes.

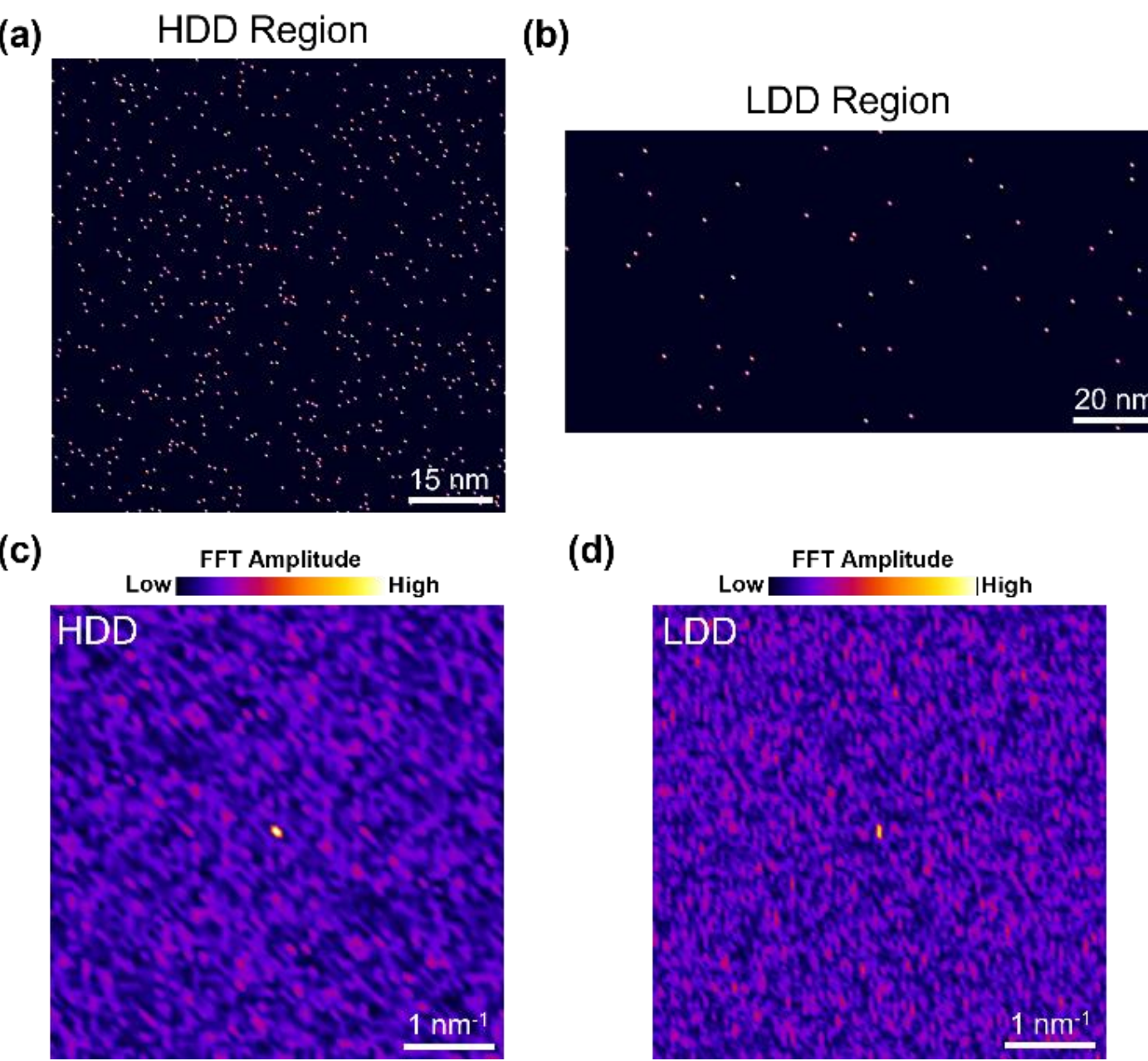

**Figure S5: Disorder distribution in HDD and LDD regimes. a,b,** Disorder distribution of (a) the HDD region and (b) LDD regions shown in the main text figures. The bright spots mark the defect locations. **c,d,** FFT amplitude maps of the disorder distributions shown in (a) and (b).

### S4. In-gap tunnel current mapping

To minimize the STM tip perturbation on 2D electron Wigner solids (WSs), we used a recently developed in-gap tunnel current measurement technique[5-7] to map the electron distribution for 2D electron WSs. As sketched in Fig. S6a, we set the STM tip Fermi level within the BL-$MoSe_2$ gap and adjust the tip-sample bias voltage ($V_S$) to match the work function difference between STM tip and BL-$MoSe_2$. This causes the electric field between the STM tip and BL-$MoSe_2$ to be minimized, thus reducing the tip perturbation on 2D electron WSs. Figure

S6b shows an example of spatially resolved tunnel current spectroscopy $|I(V_S, d)|$. The yellow dashed lines mark the edges of the BL-$MoSe_2$ conduction band and valence band where "direct" tunnel current flows as in conventional STM. Inside the BL-$MoSe_2$ gap, we notice a smaller tunnel current which is the in-gap tunnel current. Figure S6c shows a zoomed-in view of the blue dashed box in Fig. S6b where peak features corresponding to localized WS electrons can be well observed.

To acquire a 2D map of an electron WS we use "two-pass" scanning. For the first pass we select a large $V_S$ value that is outside the BL-$MoSe_2$ band gap to record the surface topography using constant-current scan conditions. For the second pass we open the feed-back loop and force the tip to follow the recorded topography of the first pass while setting $V_S$ to the value that gives the least perturbed WS images. In general, an STM tip with more negative $V_S$ will attract electrons in BL-$MoSe_2$ and an STM tip with more positive $V_S$ will repel electrons. The experimental $V_S$ that gives best WS images is typically reached near the onset of in-gap tunneling current (as sketched by the red dashed line in Fig. S6c). Near this bias range the in-gap tunneling current image will show well-localized electrons deep in the Wigner solid regime (Fig. S7a). However, if $V_S$ is too negative then the STM tip will become too attractive, thus leading to images with a "melted" appearance at that same gate voltage (Fig. S7b).

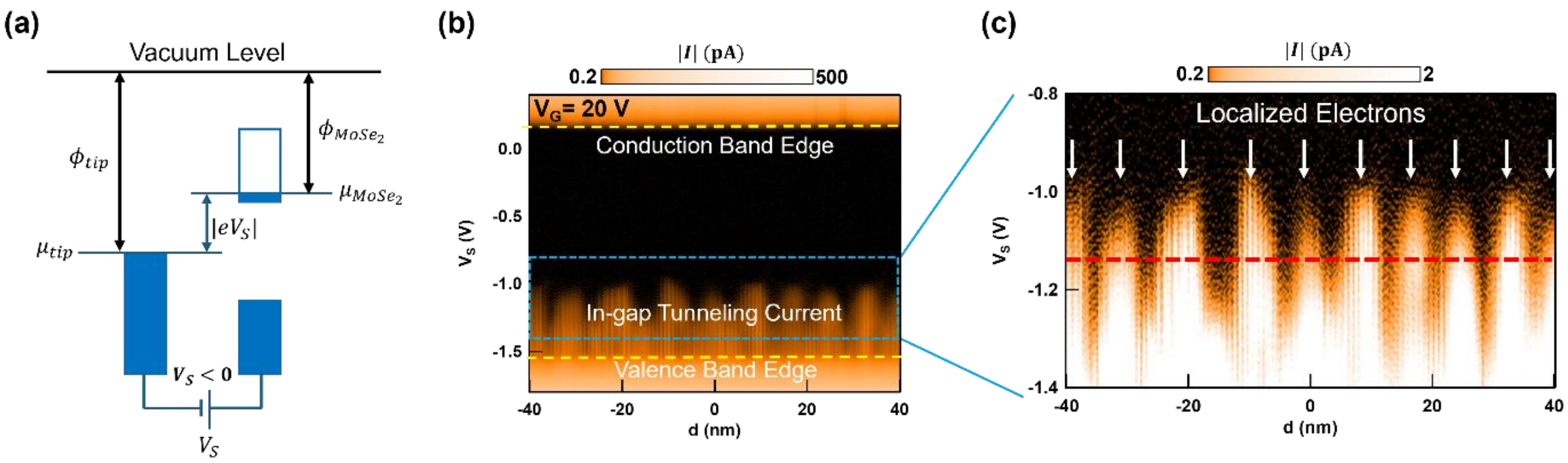

**Figure S6: In-gap tunnel current mapping. a,** Energy diagram of the in-gap tunneling technique. The tip's chemical potential is maintained in the BL-$MoSe_2$ band gap, and $V_S$ is adjusted to match the work function difference between the STM tip and the electron-doped BL-$MoSe_2$. **b,** $I(V_S, d)$ measured at $V_G = 20$ V along a line in the HDD BL-$MoSe_2$ region shown in Fig. 5a. The set point used to take the measurement was $I = 0.1$ nA, $V_S = -1.82$ V. The yellow dashed lines indicate the band edge of the BL-$MoSe_2$ conduction and valence bands. **c,** Zoomed-in view of the blued dashed line box shown in Fig. S2b. The white arrows indicate the positions of localized WS electrons. The red dashed line indicates the $V_S$ value used to take the in-gap tunnel current maps shown in Figs. 5b-e.

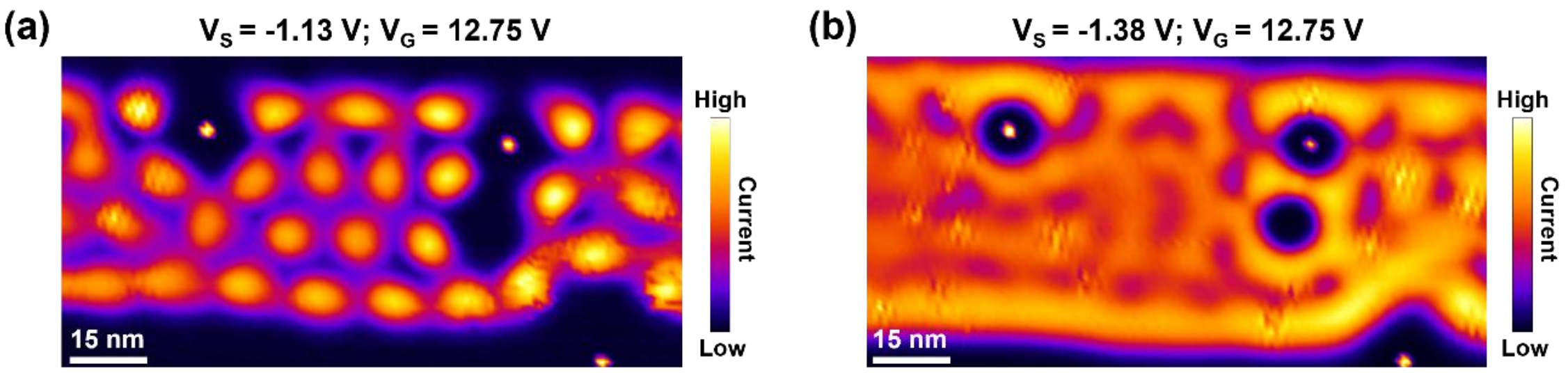

**Figure S7: $V_S$ dependence of in-gap tunneling current maps.** Experimental in-gap tunnel current maps of an LDD region measured at $V_G = 12.75$ V with (a) $V_S = -1.00$ V, and (b) $V_S = -1.38$ V.

### S5. Electron density estimation

We estimated the electron density $n_e$ at different $V_G$ values through tracking the WS wavevector and using it to count the electron number, and also combining this with device capacitance. Figure S8a-c show the squared structure factor $|s(k)|^2$ of the in-gap tunnel current maps shown in Figs. 2a,d and Fig. 3b, respectively. Clear WS $|s(k)|^2$ peaks can be observed (marked by yellow circles). The averaged WS peak positions ($k_{WS}$) are shown in the corresponding images. We convert $k_{WS}$ to $n_e$ *via* $n_e = \frac{\sqrt{3}k_{WS}^2}{8\pi^2}$ for a triangular lattice. The $n_e$ of Figs. 2b-c can be estimated by assuming a linear dependence between $n_e$ and $V_G$ when $1.6\text{ V} \leq V_G \leq 3.3\text{ V}$. For $V_G \geq 3.3$ V we used the geometric capacitance of our LDD device and the $n_e$ value estimated through $k_{WS}$ at $V_G = 3.3$ V as a reference point to estimate $n_e$. This can be explicitly expressed as $n_e(V_G) \approx (7.26 \times 10^{10}\text{cm}^{-2}/V) * (V_G - 3.3\text{ V}) + 4 \times 10^{11}\text{ cm}^{-2}$. To verify the validity of this method we compared the $n_e$ values estimated through both geometric capacitance and $k_{WS}$ at $V_G = 5.8$ V. The $n_e$ values estimated through geometric capacitance and $k_{WS}$ are $\sim 5.82 \times 10^{11}$ cm$^{-2}$ and $\sim 5.86 \times 10^{11}$ cm$^{-2}$, which are reasonably close.

When the electron WS phase is robust without melted regions, then the $n_e$ value can also be estimated by directly counting the total electron numbers ($N_e$) and charged defect numbers ($N_{CD}$). Then $n_e \approx \frac{N_e + N_{CD}}{S}$, where $S$ is the area of the measurement window. For example, Figures S9a,b show the corresponding in-gap tunneling current images shown in Figs. 2a,d with electron locations marked by blue dots and $N_e$ labeled on top. There are in total 16 charged defects in this area, but since some charged defects are very close to each other, we estimated the effective $N_{CD}$ to be $\sim 14$. Since the measurement window size is 140 nm $\times$ 70 nm, $n_e$ can be estimated to be $\sim 2.96 \times 10^{11}$ cm$^{-2}$ and $\sim 4.08 \times 10^{11}$ cm$^{-2}$ at $V_G = 1.6$ V and $V_G = 3.3$ V, respectively. These values are consistent with the $n_e$ values extracted from $k_{WS}$.

For HDD regions, we estimated $n_e$ by counting $N_e$ and $N_{CD}$ in the measurement area for Figs. 5b-d similar to the LDD region. Examples of the in-gap tunnel current images with electron locations marked by blue dots and $N_e$ labeled on top are shown in Figures S9c,d. Since the window size is 80 nm $\times$ 80 nm and $N_{CD} = 4$ for this region we estimate $n_e \approx 4.38 \times 10^{11}$ cm$^{-2}$ and $8.13 \times 10^{11}$ cm$^{-2}$ at $V_G = 10$ V and 15 V, respectively. For Fig. 5e $n_e$ is estimated through the geometric capacitance of our device using the extracted $n_e$ value of Fig. 5d as a reference point since the electron WS is partially melted in Fig. 5e.

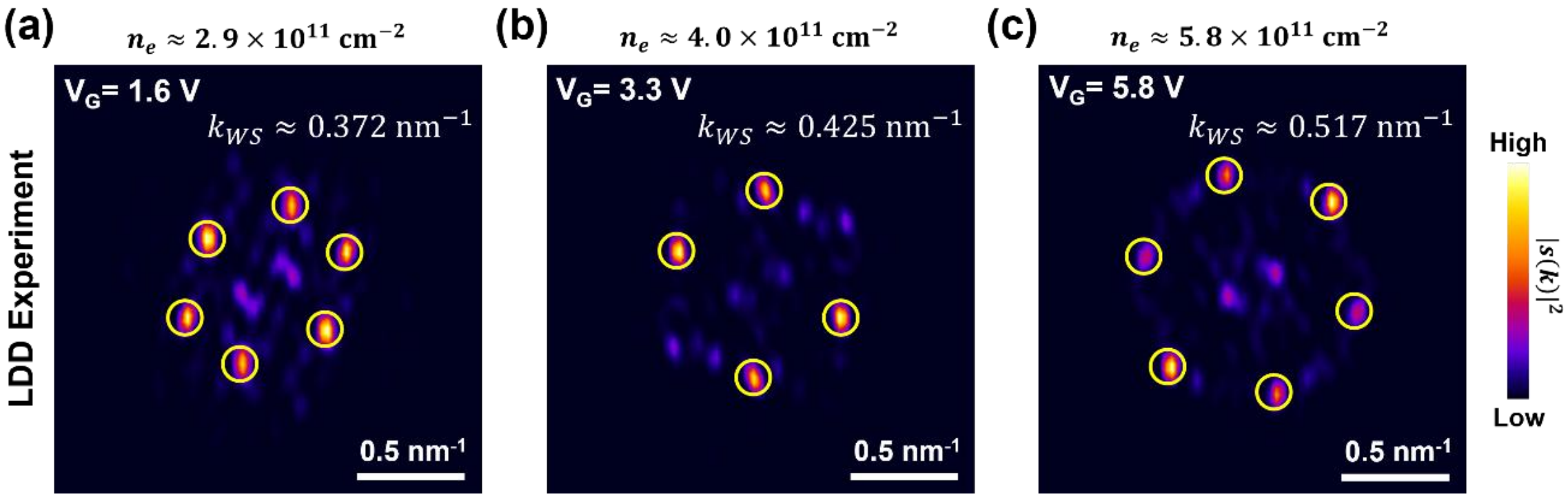

**Figure S8: Extracting electron Wigner solid wave vectors. a-c,** Squared structure factors $|s(k)|^2$ of electron WS images measured at $V_G = 1.6$ V, 3.3 V and 5.8 V for the LDD device. The yellow circles indicate the positions of the WS $|s(k)|^2$ peaks.

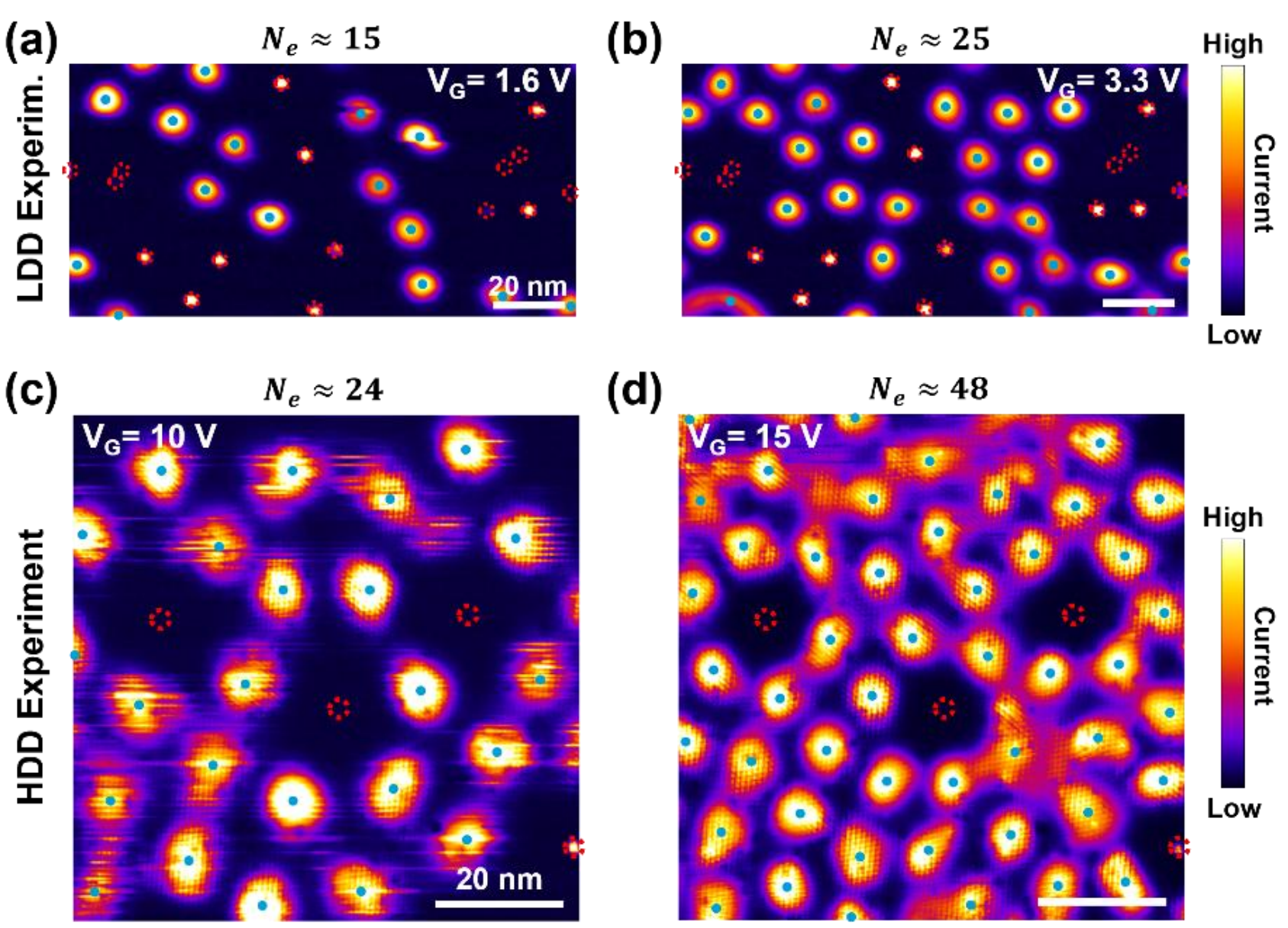

**Figure S9: Counting Wigner solid electron numbers. a-d,** The same in-gap tunnel current maps as shown in Figs. 2a,d and Figs. 5b,c. The blue dots mark the locations of WS electrons. The red dashed circles mark the locations of charged defects.

## S6. $r_s$ value estimation and impact of GNR contact

We are not able to directly measure $\varepsilon$ in our experiments and it is difficult to precisely estimate due to the complicated dielectric environment of our sample (for example, the STM tip and nearby contacts are also sources of screening). As a result, we set $\varepsilon$ as the main fitting parameter in our QMC simulations and optimize it through comparison of STM and QMC simulation results (see SI section S14 and Fig. S21). Our QMC results show that the main behavior of the Wigner solids we observe (e.g. electron positions, local re-entrant melting/crystallization) do not sensitively depend on the choice of $\varepsilon$ values, although it does slightly alter the overall extent of quantum melting. We experimentally investigated the influence of the GNR contact on electronic behavior in BL-$MoSe_2$ and observed that it can cause band-bending in regions very close ($\lesssim$ 10 nm) to the GNR contact edge, including depletion of electrons very near the contact (Fig. S10). But this influence is screened at far enough distance ($\gtrsim$ 15 nm). The experimental data shown in our main text were acquired at regions far enough away from the GNR contacts that their influence is negligible.

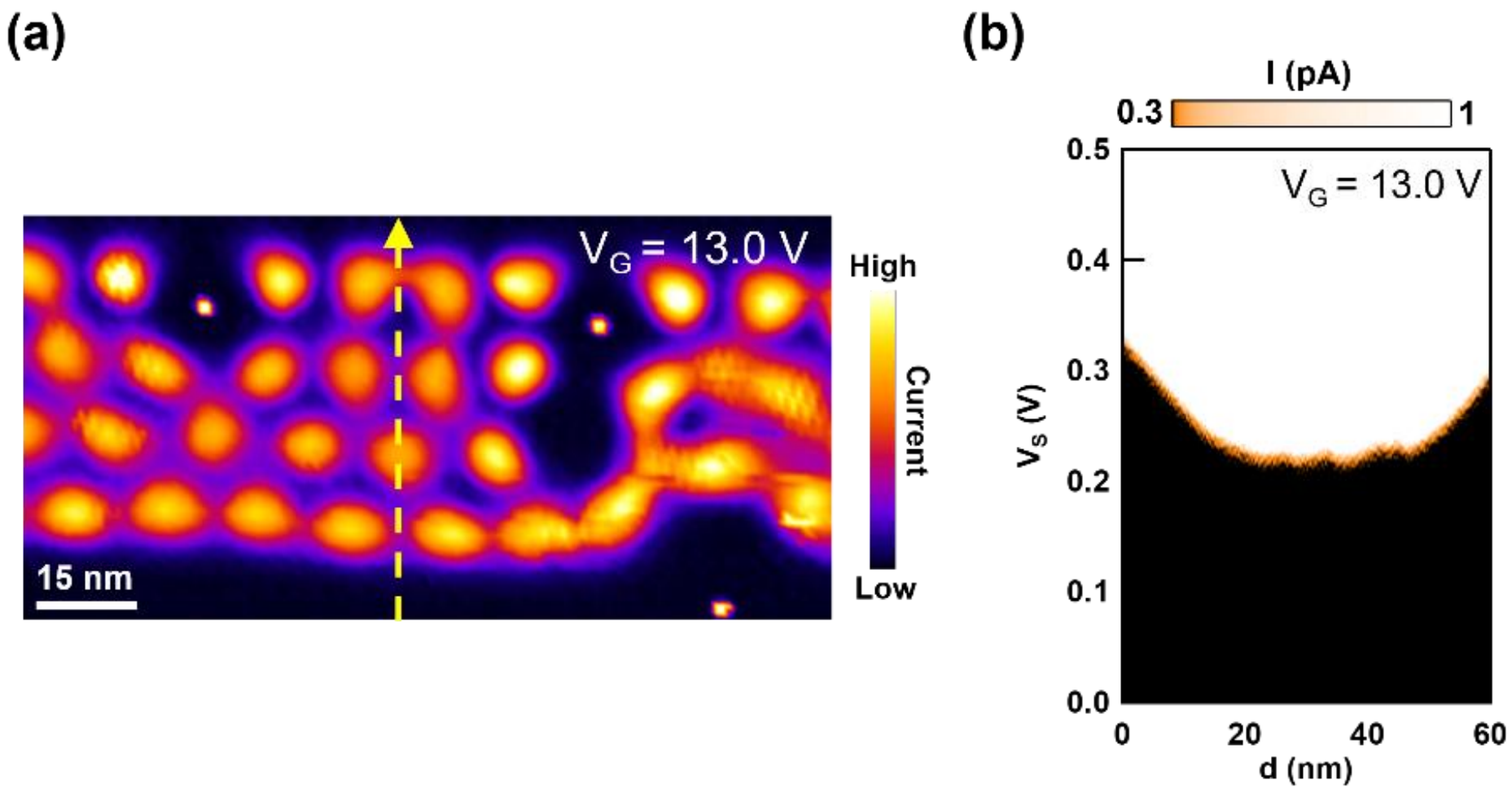

**Figure S10: Influence of graphite nanoribbon contact on electron Wigner solids. a.** In-gap tunneling current image measured for a BL-$MoSe_2$ region where the top and bottom boundaries are very close to GNR contacts (~10 nm away). The image was measured at $V_G = 13$ V, $V_S = -1.14$ V. Electrons avoid the top and bottom boundaries close to the GNR contacts. **b,** Color plot of $I(V_S, d)$ measured across the dashed yellow arrow in (a). Setpoint: $I = 2$ nA, $V_S = 1.36$ V, $V_G = 13$ V. The conduction band edge bends toward more positive $V_S$ at locations closer to the GNR contacts.

### S7. Temperature and $n_e$-dependent transport data for LDD BL-$MoSe_2$ Device

We have performed temperature and density-dependent transport measurements for electron-doped BL-$MoSe_2$ (the same material as used in our STM measurements) (Fig. S11). The BL-$MoSe_2$ used in our transport device was exfoliated from the same parent crystal source as the STM LDD device shown in our main text. Figures S11a-p show the one-to-one correlation between LDD STM images and temperature dependent resistance measurements for BL-$MoSe_2$ in the density range $2.9 \times 10^{11}$ cm$^{-2}$ $\leq n_e \leq 2.55 \times 10^{12}$ cm$^{-2}$. Good agreement is found between the STM-visualized melting of an electron Wigner solid and the metal-insulator transition observed via transport. At our STM measurement temperature (~4.8 K) the transport data show clear insulator behavior ($\frac{dR}{dT} < 0$) for BL-$MoSe_2$ with $2.9 \times 10^{11}$ cm$^{-2}$ $\leq n_e \leq$ $6.3 \times 10^{11}$ cm$^{-2}$ (Figs. S11a-g), a metal-insulator transition ($\frac{dR}{dT}$~0) at $n_e \approx 7.3 \times 10^{11}$ cm$^{-2}$ (Fig. S11h), and metallic behavior ($\frac{dR}{dT} > 0$) for for $8.3 \times 10^{11}$ cm$^{-2}$ $\leq n_e \leq 2.54 \times$ $10^{12}$ cm$^{-2}$(Figs. S11i-p). This corresponds nicely to our STM data which reveals a solid electron phase for BL-$MoSe_2$ with $2.9 \times 10^{11}$ cm$^{-2}$ $\leq n_e \leq 7.2 \times 10^{11}$ cm$^{-2}$ (Figs. S11a-h), and an onset of electron Wigner solid melting at $n_e \approx 8.1 \times 10^{11}$ cm$^{-2}$ (Fig. S11i). Thus, in addition to the strong support from our NQS-QMC simulations, our transport data further confirms our experimental observation of electron Wigner solids in the presence of quenched disorder and their quantum melting behavior. This is the first combined STM/Transport study ever performed in this physical regime.

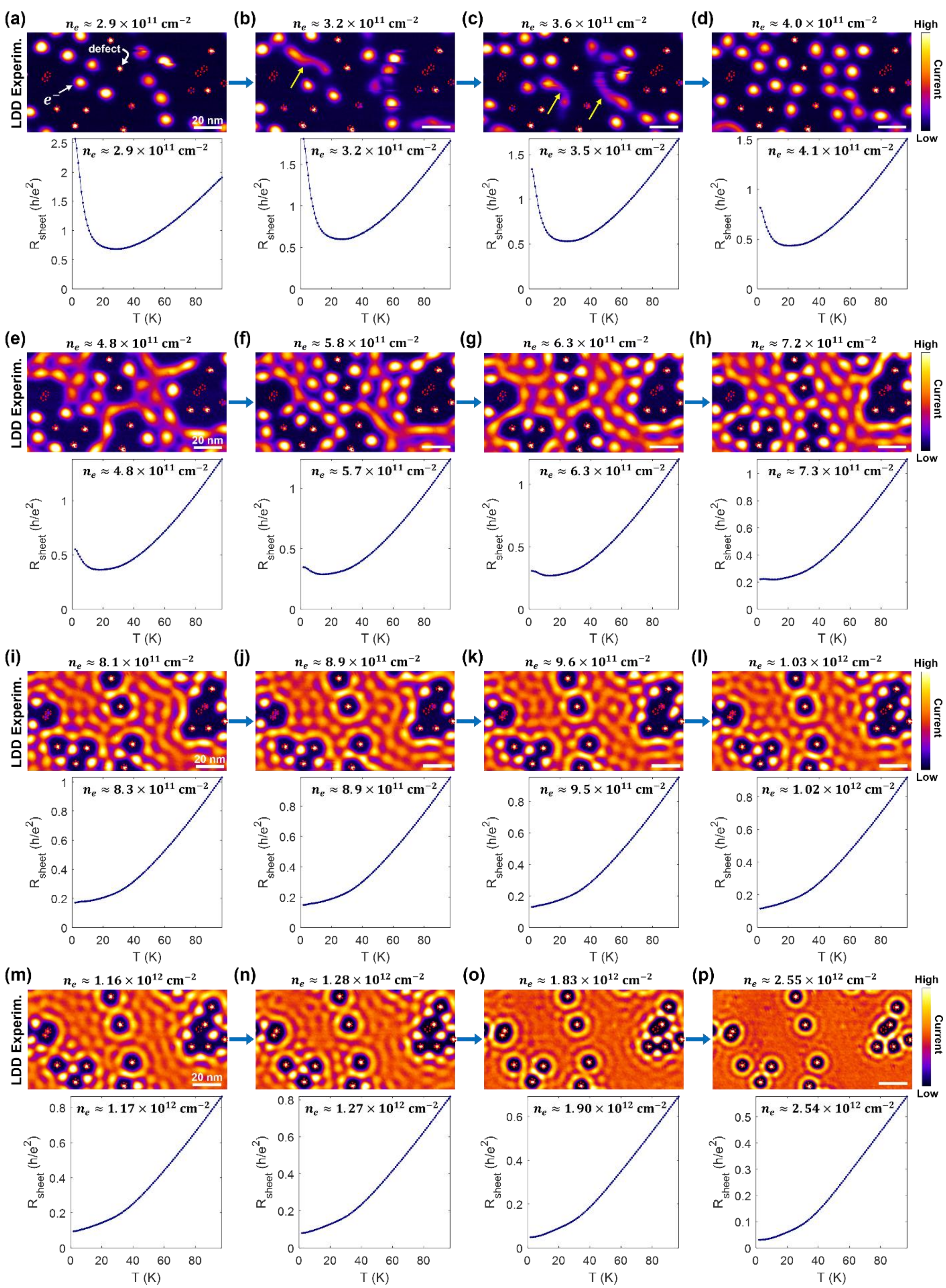

**Figure S11: Correlation between STM-visualized electron Wigner solid melting and metal-insulator transition in 2D transport. a-p,** In-gap tunneling current images are shown for a LDD BL-$MoSe_2$ device (same as in the main text Figs. 2-4) for $2.9 \times 10^{11}$ cm$^{-2}$ $\leq n_e \leq$ $2.55 \times 10^{12}$ cm$^{-2}$. The corresponding temperature-dependent resistivity curves for a similar LDD BL-$MoSe_2$ device at nearly identical $n_e$ values are displayed below each STM image. The STM-visualized melting of the electron Wigner solid as well as the transport-observed metal-insulator transition both occur at $n_e \sim 7.5 \times 10^{11}$ cm$^{-2}$.

## S8. Experimental and simulated radial averaged electron density line cuts around isolated charged defects

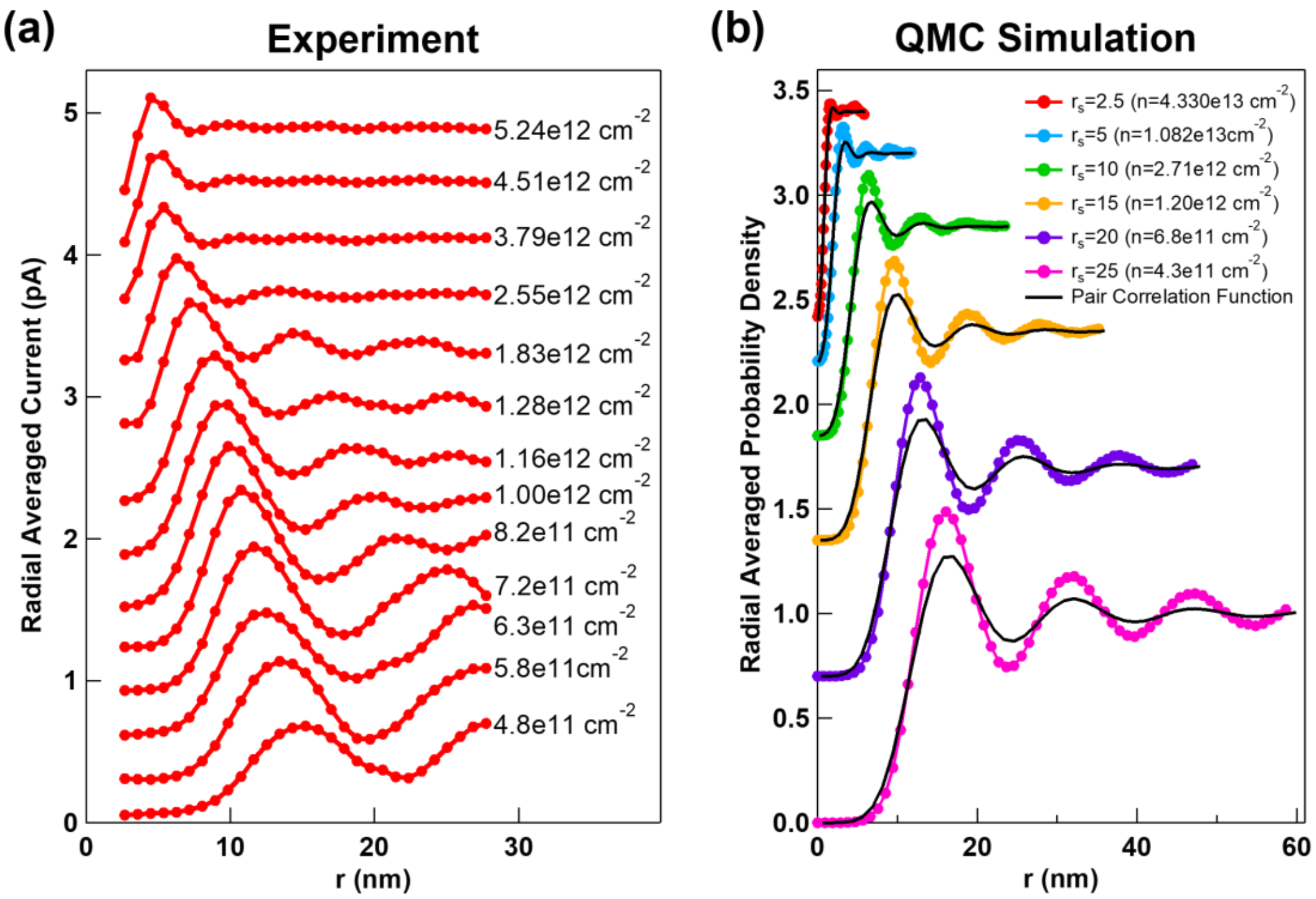

**Figure S12: Experimental and simulated radial averaged electron density line cuts around an isolated charged defect. a,** Radial averaged line cuts of the in-gap tunnel current intensity around the charged defect marked by the green dashed box in Fig. 4c. Line cuts are shown for different values of $n_e$. **b,** The colored lines are QMC simulated radial averaged electron probability density line cuts around a single charged defect at different $n_e$. The overlaid black lines show the calculated pair correlation function at each $n_e$ value. The dielectric constant used in the QMC simulations is $\varepsilon = 3.73\varepsilon_0$.

**S9. Comparing experimental charge density oscillations around charged defects with characteristic Wigner crystal and 2D $\boldsymbol{k_F}$ length scales**

To further analyze the Friedel oscillation behavior in BL-$MoSe_2$ we extracted the distances from the defect center to the first peak ($a_1$) and from the first to the second peak ($a_2$) in both experimental and QMC simulated Friedel oscillation line profiles shown in Fig. S12 for different $n_e$ values and plotted them as a function of $1/\sqrt{n_e}$ in Fig. S13. We then compare the $n_e$ dependence of $a_1$ and $a_2$ with the theoretical $n_e$ dependence of expected peak spacing for a radially averaged Wigner crystal as well as the probability density of a non-interacting 2D wavefunction at $k_F$. For an ideal Wigner crystal with a triangular lattice, the first peak position of its radial distribution function equals its lattice constant. Thus, the first peak distance in a radial averaged probability density line profile for a Wigner crystal at $n_e$ can be expressed as:

$$a_{WC} = \sqrt{\frac{2}{\sqrt{3}}}\,\frac{1}{\sqrt{n_e}} \approx 1.075\frac{1}{\sqrt{n_e}}$$

For a non-interacting 2DEG we expect to see Friedel oscillations with wavevector $k_{FO} = 2k_F$. The period for non-interacting Friedel oscillations in a 2D electronic system with $\nu$-fold degeneracies at $n_e$ can then be expressed as:

$$a = \frac{2\pi}{k_{FO}} = \frac{\sqrt{\nu\pi}}{2}\frac{1}{\sqrt{n_e}}$$

The peak spacing for a radial averaged probability density line cut for non-interacting Friedel oscillations in a single-valley system with 2-fold degeneracy can then be expressed as:

$$a_{single-valley\ FO} = \frac{\sqrt{2\pi}}{2}\frac{1}{\sqrt{n_e}} \approx 1.253\frac{1}{\sqrt{n_e}}$$

For BL-$MoSe_2$, the electrons are located at six Q valleys[8], and so we also considered radial averaged line cut peak spacing for a six-valley system (i.e., 12-fold degeneracy) which can be expressed as:

$$a_{six-valley\ FO} = \sqrt{3\pi}\frac{1}{\sqrt{n_e}} \approx 3.070\frac{1}{\sqrt{n_e}}$$

We observe that $a_1$ and $a_2$ extracted from experiment and QMC simulations both follow the trend expected for the electron-electron separation in a 2D Wigner crystal (orange line) and significantly deviate from the peak-to-peak spacing expected for Friedel oscillations in a non-interacting single-valley (blue line) and six-valley (green line) 2D Fermi liquid. These results suggest that electron-electron interactions significantly modify short-range Friedel oscillation behavior in the LDD regime compared to non-interacting 2D systems.

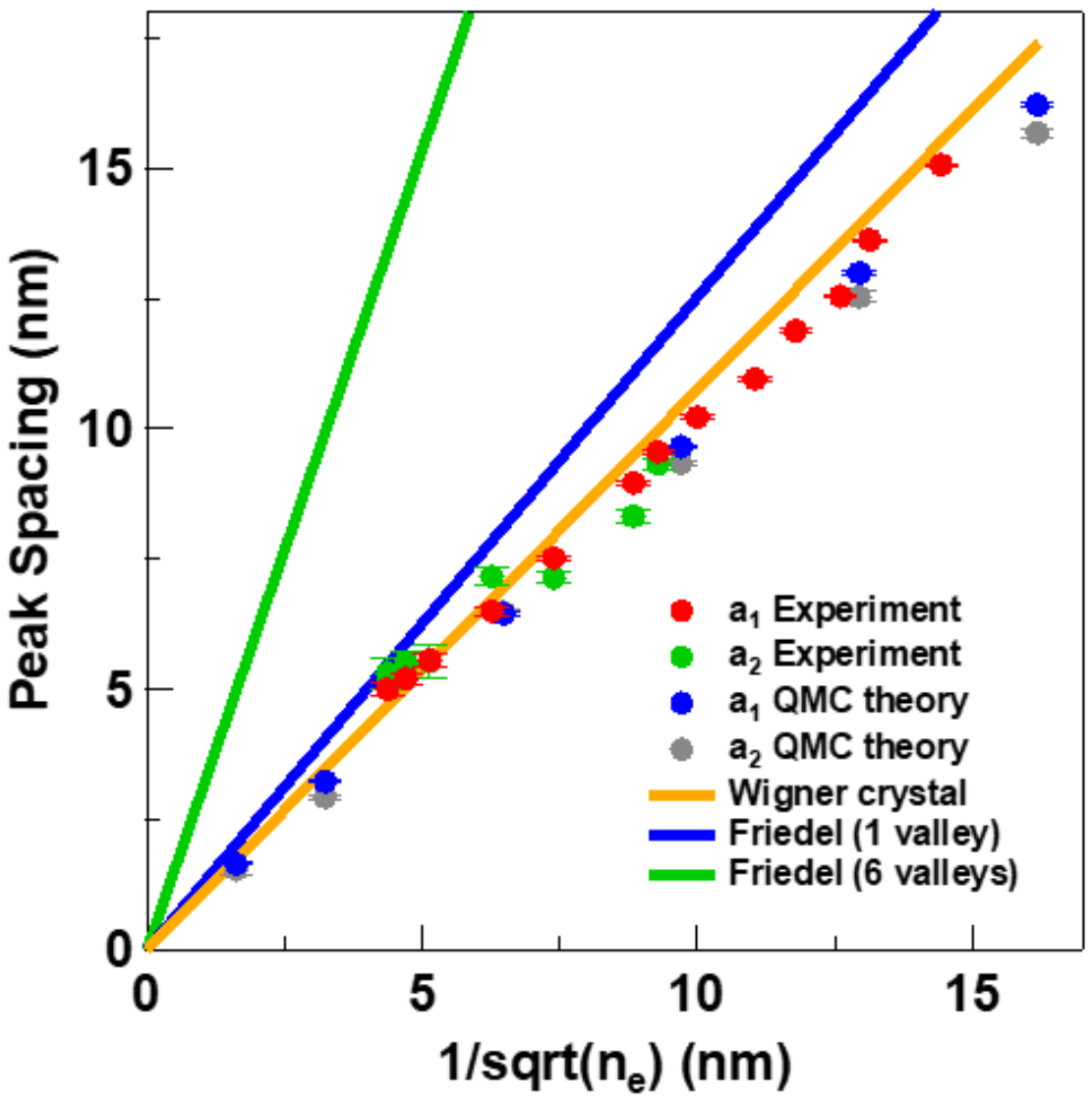

**Figure S13: $n_e$ dependence of Friedel oscillation peak spacing.** Friedel oscillation peak spacings as a function of $1/\sqrt{n_e}$ extracted from experimental and QMC simulated radial averaged density profiles around the isolated charged defect marked in Fig. 4c. $a_1$ and $a_2$ represent the spacing between the first peak and defect center (for $a_1$) and between the second peak and first peak (for $a_2$). The orange, blue and green lines represent the theoretical peak spacing of radial averaged line profiles for a Wigner crystal and non-interacting Friedel oscillations in a single-valley and six-valley system, respectively.

## S10. Non-interacting Friedel oscillation line profile calculation method

The general charge density oscillation $\delta n(r)$ near a defect potential with the form $V(r) = A/r^{\eta}$ in a non-interacting 2D electronic system derived by linear response theory can be expressed as[9]:

$$\delta n(r) = \frac{mk_F^{\eta}A}{\pi}\frac{1}{(k_F r)^{\eta}}\left[{}_1F_2\left(-\frac{1}{2};1-\frac{\eta}{2},1-\frac{\eta}{2};-(k_F r)^2\right) - \frac{\pi^{\frac{5}{2}}(k_F r)^{\eta}}{\eta\Gamma\left(\frac{3-\eta}{2}\right)\left[\Gamma\left(\frac{\eta}{2}\right)\right]^3 \sin^2(\frac{\pi\eta}{2})}\,{}_1F_2\left(\frac{\eta-1}{2};1,1+\frac{\eta}{2};-(k_F r)^2\right)\right],$$

where $m$ is the effective electron mass, $k_F$ is the Fermi wavevector, ${}_1F_2(a;b,c;z)$ is the generalized hypergeometric function, and $\Gamma(z)$ is the gamma function. We set $\eta = 1$ and use $n_e$ extracted from experiment to determine $k_F$, which is related as:

$$k_F = \sqrt{\frac{4\pi n_e}{g_v g_s}},$$

where $g_v$ and $g_s$ represent the valley and spin degeneracy of the 2D electronic system. For the non-interacting Friedel oscillation line profile shown in Fig. 4i, we set $n_e = 1.83 \times 10^{12}\ \mathrm{cm}^{-2}$, $g_v = 1$, and $g_s = 2$. We scale and shift the resulting $\delta n(r)$ in the y-axis direction to reasonably compare it with experiment in Fig. 4i. Since the BL-$MoSe_2$ conduction band has six valleys, we have also calculated the non-interacting Friedel oscillation $\delta n(r)$ for $n_e = 1.83 \times 10^{12}\ \mathrm{cm}^{-2}$, $g_v = 6$, and $g_s = 2$. This result is shown in Fig. S14 and we see that the deviation between experiment and the non-interacting Friedel oscillation is even more prominent.

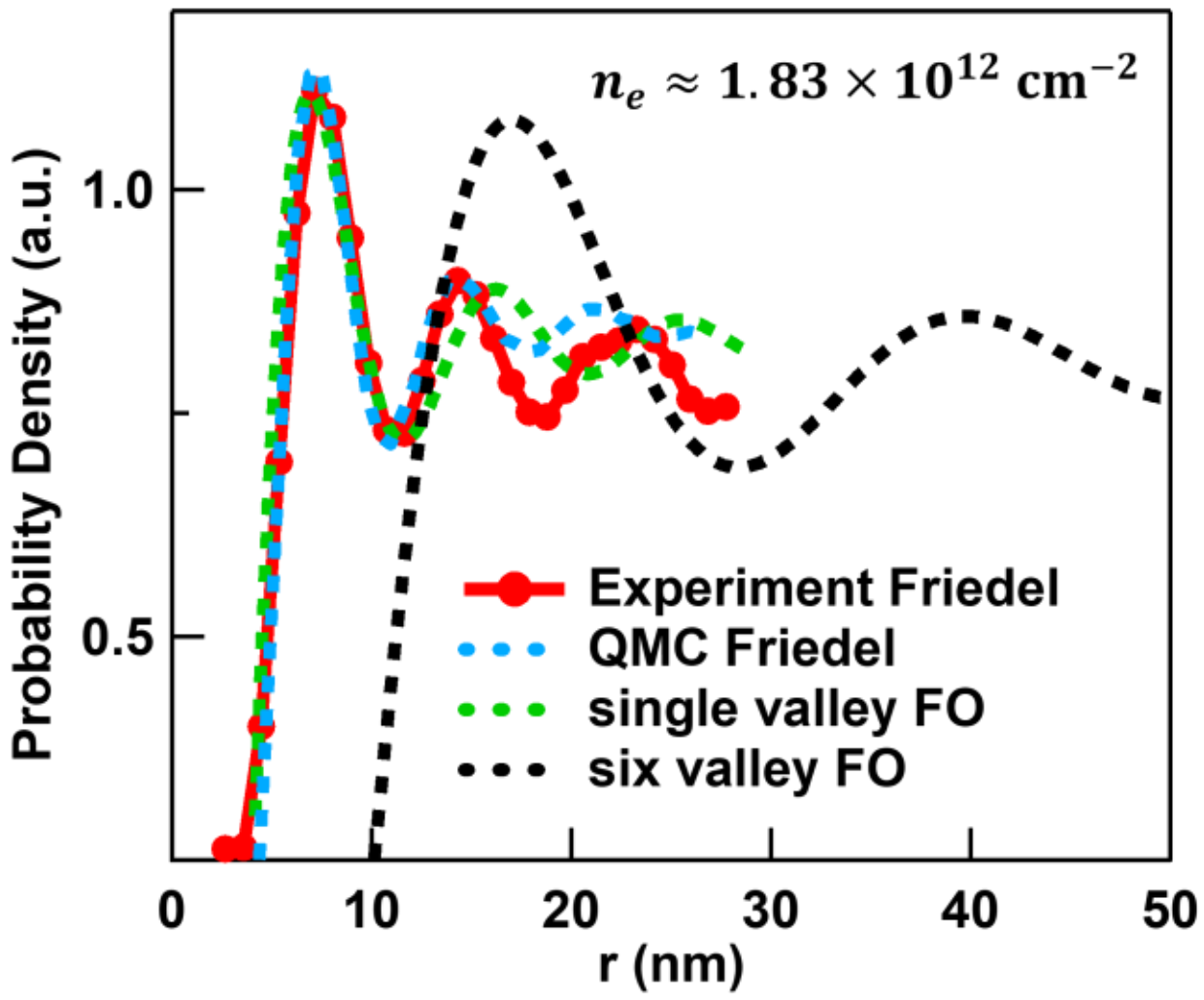

**Figure S14: Non-interacting Friedel oscillations in six-valley 2D electronic system.** The experimentally extracted (red line) and QMC simulated (blue line) Friedel oscillation line cuts together with calculated non-interacting Friedel oscillations for single-valley (green line) and six-valley (black line) 2D electronic systems at $n_e \approx 1.83 \times 10^{12}$ cm$^{-2}$.

### S11. Disappearance of mesoscopic crystalline order in HDD devices

As shown in Fig. S15, the WS phase in the HDD regime does not show isolated s(k) peaks on the mesoscopic scale, unlike the LDD regime. Instead it has a ring-like feature in s(k) and so the system lacks long-range translational order. We can extract the radius of this ring-like feature ($k_{HDD}$) as depicted in Fig. S16a by gaussian peak fitting for a radial averaged s(k) line plot (Fig. S16b). We used this process to extract $k_{HDD}$ as a function of $V_G$ and we find that it follows the expected behavior of an ideal Wigner crystal. To see this we convert $k_{HDD}$ to $n_e$ using the relation between $n_e$ and $k$ for an ideal Wigner crystal:

$$n_e = \frac{\sqrt{3}k_{HDD}^2}{8\pi^2}.$$

The results are shown by red dots in Fig. S16c. We find that the slope of $n_e$ vs $V_G$ found in this way agrees well with $n_e$ determined from $V_G$ using geometric capacitance of our 2D device (red solid line in Fig. S16c). There is a constant offset ($\sim 3.5 \times 10^{11}$ cm$^{-2}$) between the $n_e$ found from $k_{HDD}$ and the $n_e$ calculated from $V_G$ using geometric capacitance. We attribute this offset to a background doping shift and/or residual tip-gating due to non-perfect tip-sample workfunction matching. After compensating for the offset (blue dots in Fig. S16c), the $n_e$ found from $k_{HDD}$ matches very well with $n_e$ determined from geometric capacitance for $V_G < \sim 40$ V. At higher $V_G$ the deviation between the capacitively determined $n_e$ and $n_e$ determined from $k_{HDD}$ grows quickly. $V_G \sim 40$ V might mark a transition point between different physical regimes, such as when the system goes from a Coulomb-interaction-dominated Wigner solid regime (at lower $V_G$) to a disorder-dominated-Anderson-localization-regime (at higher $V_G$). Confirming this interpretation is beyond the scope of our manuscript.

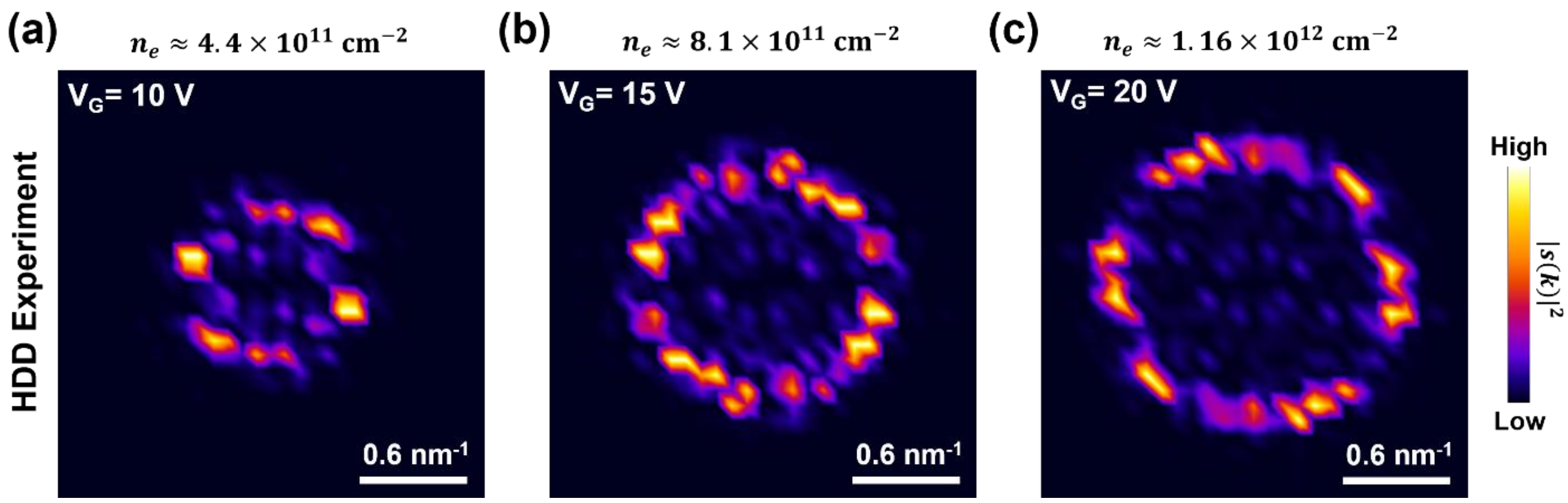

**Figure S15: Disappearance of mesoscopic crystalline order in HDD devices. a-c,** Squared structure factors $|s(k)|^2$ of electron WS images measured at $V_G = 10$ V, 15 V, and 20 V for the HDD device shown in Figs. 5b-d. Isolated $|s(k)|^2$ peaks are replaced by a diffuse ring, indicating the disappearance of mesoscopic crystalline order in the HDD regime.

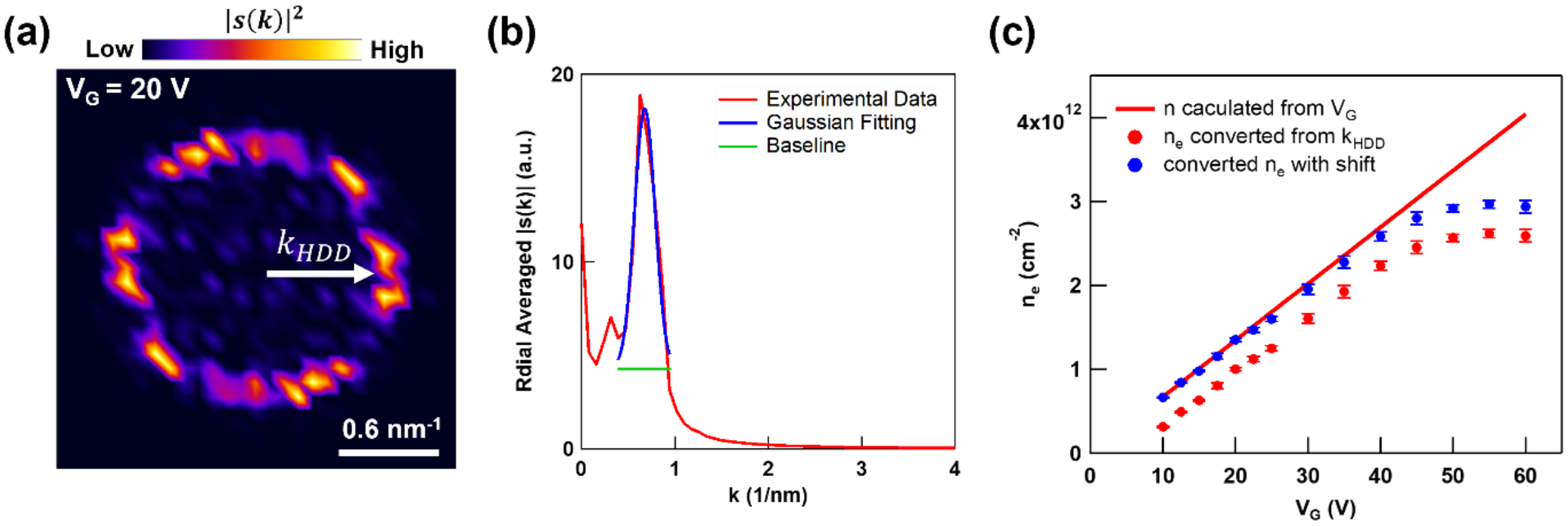

**Figure S16: Gate-dependence of the ring-like $s(k)$ feature for WSs in the HDD regime. a,** $|s(k)|^2$ for the disordered WS shown in Fig. 5d. The radius ($k_{HDD}$) of the ring-like feature is marked by a white arrow. **b,** Radial-averaged $s(k)$ for the WS shown in Fig. 5d (red line). A Gaussian fit (blue line) is used to extract $k_{HDD}$. The green line is the baseline used for the Gaussian fit. **c,** Comparison between $n_e$ calculated from $V_G$ using device geometric capacitance (red solid line), $n_e$ obtained from $k_{HDD}$ (red dots), and $n_e$ obtained from $k_{HDD}$ plus a $3.5 \times 10^{11}$ cm$^{-2}$ offset (presumably a doping shift).

**S12. Bond orientational order analysis for WSs**

We examined the bond-orientational (BO) order of the WS phase in the HDD regime. We evaluated the BO order parameter[10] for n-fold rotational symmetry:

$$\psi_n(r_i) = \frac{1}{N_i} \sum_{k=1}^{N_i} e^{in\theta_{ik}},$$

where $r_i$ is the location of the i$^{th}$ electron, $N_i$ is the number of bonds between the i$^{th}$ electron and its nearest neighbors (including electrons at charged defect sites) as determined from Delaunay triangulation, and $\theta_{ik}$ is the angle of the bond between the i$^{th}$ electron and k$^{th}$ neighboring electron. We calculated $\psi_n$ for 4,5,6,7,8-fold rotational symmetries and determined the primal symmetry at each electron site by finding the n-fold symmetry that gives the maximum $|\psi_n|$. Figures S17a-c and Figs. S17d-f show the primal local BO order map and the corresponding histogram of dominant rotational symmetry for the WSs in the HDD regime shown in Figs. 5b-d, respectively. We observe that the local BO order for the disordered WS states at each $n_e$ value in the HDD regime has dominant 6-fold rotational symmetry, which helps to explains why $k_{HDD}$ follows the behavior of an ideal Wigner crystal.

We applied a similar primal local BO order analysis to WSs in the LDD regime of Figs. 2a-d and Figs. 3a-d. The corresponding primal local BO order maps and histograms are shown in Figs. S18 and S19. We notice that the primal local BO order for WSs in the LDD regime strongly fluctuates as the system undergoes local re-entrant melting/crystallization. The LDD WSs are dominated by 6-fold rotational symmetry when the system is in a "locked" solid state (Figs. S18a,d and Figs. S19b,d), but this dominance significantly reduces or even disappears when the system is in a local melted state (Figs. S18b,c and Figs. S19a,c).

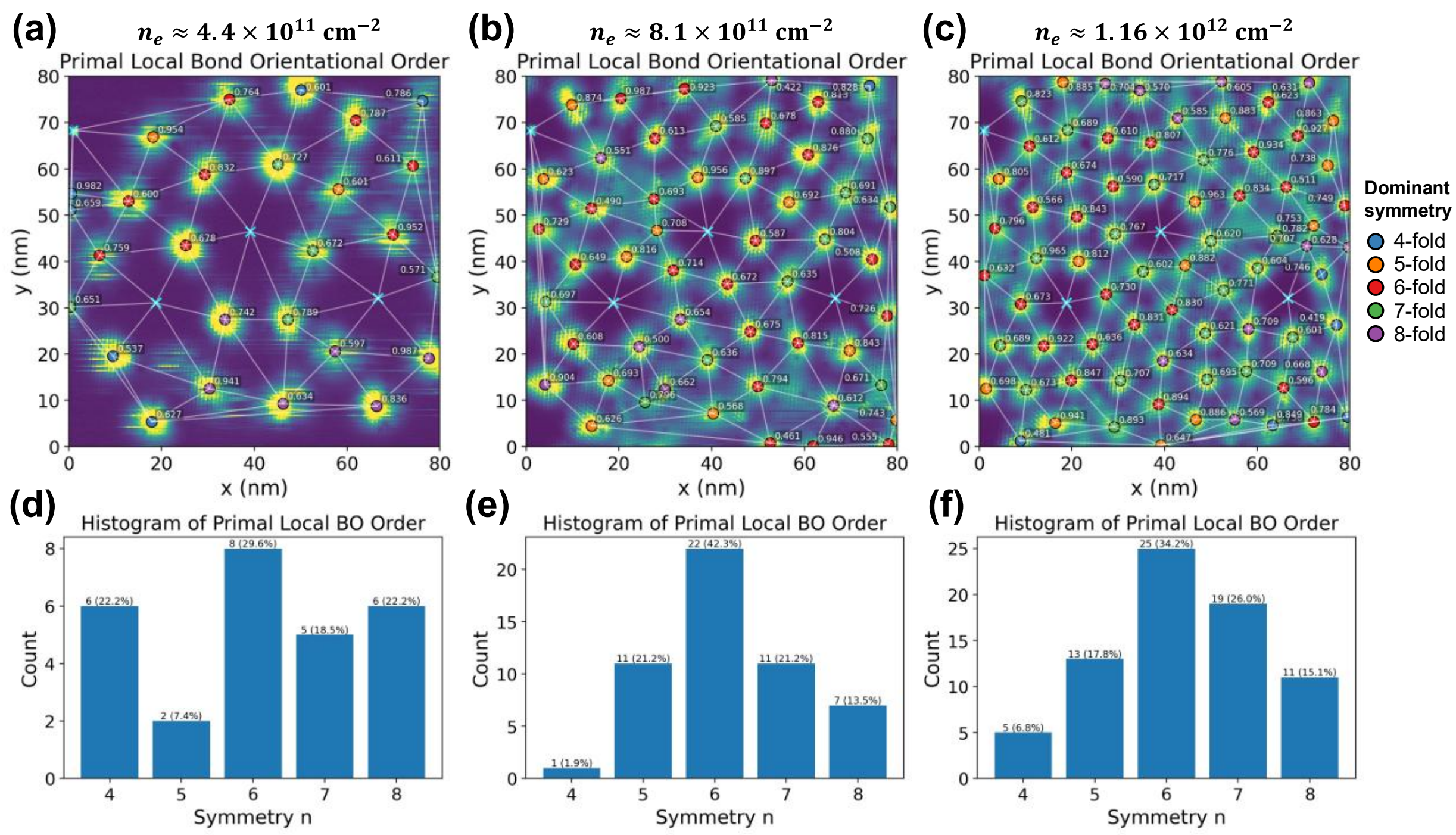

**Figure S17: Primal local bond orientational order for the WS phase in the HDD region. a-c,** Primal local bond orientational order map for WS images shown in Figs. 5b-d, respectively. The dominant rotational symmetry at each electron site is color coded. **d-f,** Histograms reveal the dominant rotational symmetry corresponding to the primal local bond orientational order maps in a-c.

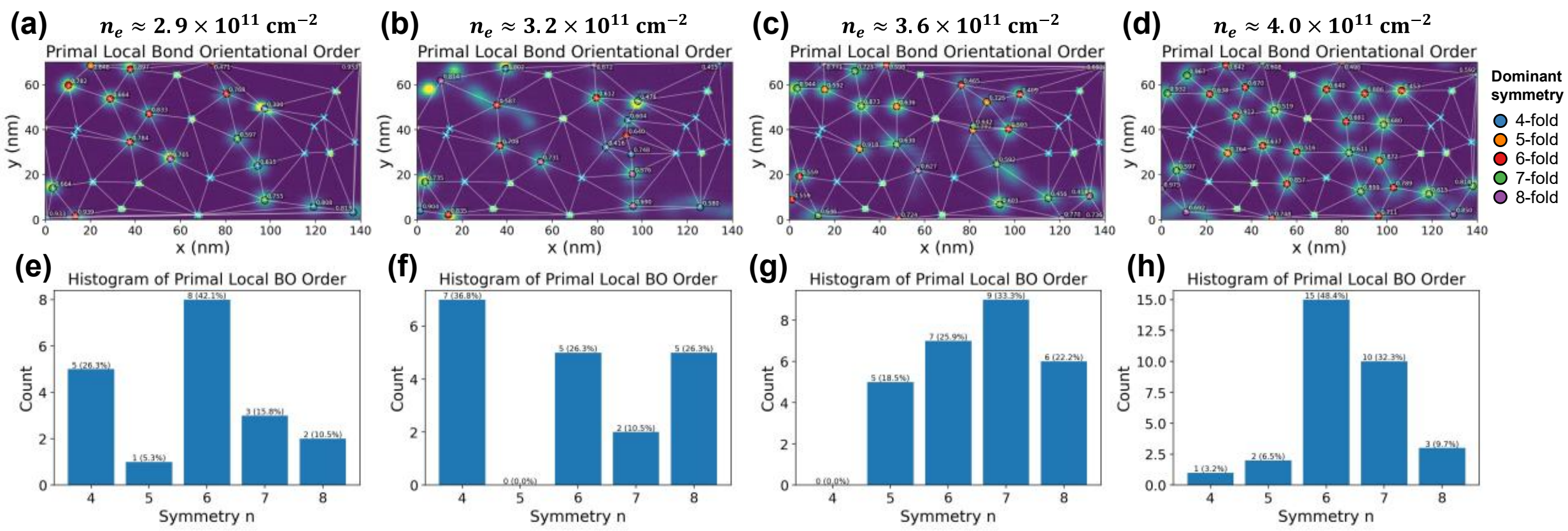

**Figure S18: Primal local bond orientational order for the WS phase in the LDD region at low $n_e$. a-d,** Primal local bond orientational order maps for WS images shown in Figs. 2a-d. The dominant rotational symmetry at each electron site is color coded by added dot. **e-h,** Histograms show the dominant rotational symmetry for the primal local bond orientational order maps in a-d.

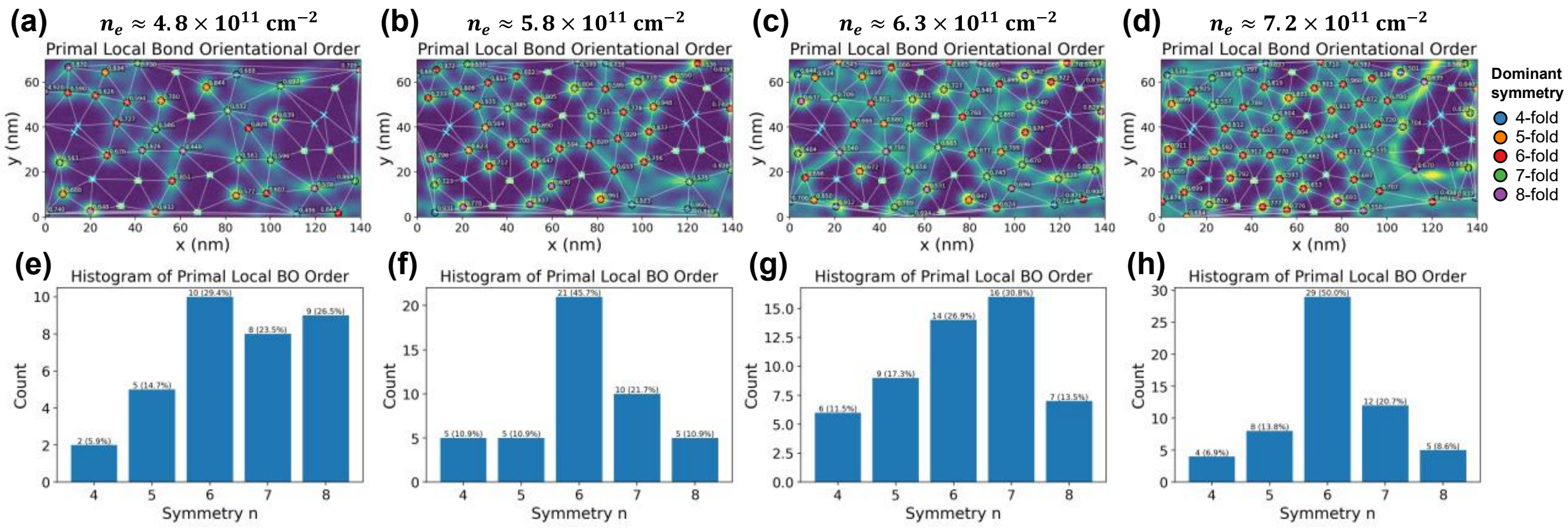

**Figure S19: Primal local bond orientational order for the WS phase in the LDD region at intermediate $n_e$. a-d,** Primal local bond orientational order maps for WS images shown in Figs. 3a-d. The dominant rotational symmetry at each electron site is color coded. **e-h,** Histograms reveal the dominant rotational symmetry corresponding to the primal local bond orientational order maps in a-d.

## S13. Quenching of Friedel oscillations at higher $n_e$ in HDD devices

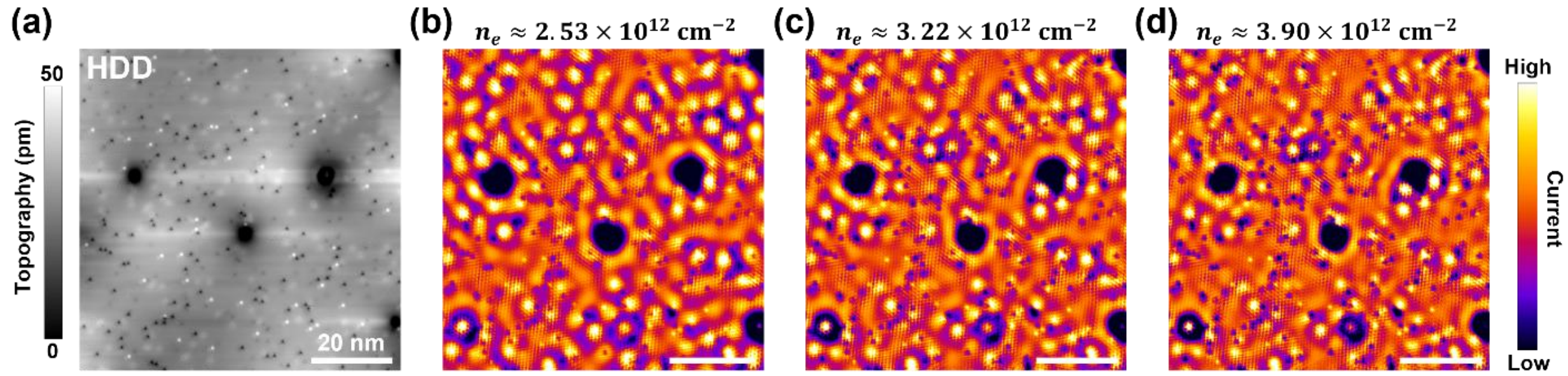

**Figure S20: Quenching of Friedel oscillations at higher $n_e$ in HDD devices. a,** STM topograph of the HDD BL-$MoSe_2$ device shown in Fig. 5a. **b-d,** Experimental in-gap tunnel current maps measured at (b) $V_G = 40.00$ V, (c) $V_G = 50.00$ V, (d) $V_G = 60.00$ V. Setpoint: $V_S = -1.14$ V. Estimated experimental $n_e$ values are shown on top of each image. The in-gap tunnel current maps remain almost unchanged when increasing $n_e$ in this range. Heterogeneity in the local electron density has a one-to-one correspondence with the isovalent defect positions in (a).

## S14. Quantum Monte Carlo (QMC) simulation method for disordered electron Wigner solids

We use variational Monte Carlo (VMC) to simulate the ground-state of a two-dimensional electron gas (2DEG) in the presence of disorder. The Hamiltonian in effective Bohr radius ($a_B^* = 4\pi\varepsilon\hbar^2/m^*e^2$) and Hartree units ($\mathrm{Ha}^* = \hbar^2/m^*{a_B^*}^2$) is given by

$$H == -\sum_i \frac{1}{2}\nabla_i^2 + \sum_{i<j} V_c(r_i - r_j) + \sum_{iJ} V_c(r_i - R_J) + \sum_{iJ} V_{sc}(r_i - s_J)$$

where $V_c$ is the unscreened coulomb interaction between electrons at positions $r$ and long-range impurities at positions $R$. The screened interaction between electrons and short-range impurities at positions $s$ is approximated as dual-gate screened Coulomb interaction given by

$$V_{sc}(|r_i - s_J|) = \frac{1}{(2\pi)^2}\int dq e^{iq\cdot(r_i - s_J)} v_{sc}(|q|), \qquad v_{sc}(|q|) = 2\pi \frac{tanh(d_g|q|)}{|q|}$$

where $d_g$ is the gate distance. We calculate this interaction only between electrons and short-range impurities.

To determine the ground state numerically, we adopt the unified $(MP)^2$-NQSs ansatz[11] which can describe both Wigner solid and electron liquid states without imposing *a priori* bias in the trial wavefunction. The ansatz is slightly modified to capture electron disorder interaction by adding an electron-defect Jastrow factor. The final ansatz also contains spin-resolved electron-electron Jastrow factors in addition to the all-electron neural-network Jastrow and backflow transformation. The electron coordinates, transformed by the neural-network backflow, are fed into orbitals made up of linear combination of planewaves. Optimization and hyperparameters for the long-range impurities are chosen to be consistent with the previous 2DEG calculations with only slight adjustments. We typically use stochastic reconfiguration (SR)[12] to optimize the ansatz, but use the ADAM optimizer[13] for several calculations with many short-range impurities because it is more numerically stable and computationally efficient.

We show in Table I the total energies of all simulations shown in the main text labeled according to figure in the rows and subfigure in the columns. We have also uploaded system parameters such as defect configurations, $r_s$ values, and screening strengths to a data repository[14] where we have also included a selection of training curves as well as a Python script to plot the densities corresponding to the figures in the main text produced with QMC.

In our QMC simulations charged defects are treated as point charges with net charge $-e$, and their interactions with electrons are evaluated using the Ewald summation technique. Isovalent defects are treated as screened charges with exponentially decaying interactions[15,16] characterized by an input "gate distance" parameter which we set to $d_g = 0.15a_0^*$ ($\sim$0.05 nm), where $a_0^*$ is the effective Bohr radius. Some isovalent defects are modeled as attractive and others as repulsive. We determined whether an isovalent defect is attractive or repulsive by experimentally observing its influence on electron probability density in the liquid regime. Some isovalent defects have a completely undetectable perturbation to the electron probability density and so we exclude those isovalent defects from our QMC simulations. This results in an effective isovalent defect density of $n_0 \approx 7.63 \times 10^{12}$ cm$^{-2}$ as described in the caption of Fig. 5.

The value of the dielectric constant $\varepsilon$ used in our QMC simulations is a fitting parameter and is selected based on comparing the similarity between experimental and QMC electron density maps (it is the only fitting parameter in our simulations). We notice that $\varepsilon = 2.58\varepsilon_0$ is a better fit compared to $\varepsilon = 3.73\varepsilon_0$ in the low $n_e$ regime. However, in the intermediate and high $n_e$ regime, $\varepsilon = 3.73\varepsilon_0$ provides a better fit. For example, in the low $n_e$ regime *experimental* electron density maps show well isolated and narrow electron wavepackets (Figs. S21a,b). QMC results obtained with $\varepsilon = 2.58\varepsilon_0$ (Figs. S21i,j) have narrower and more isolated electron wavepackets compared to those obtained with $\varepsilon = 3.73\varepsilon_0$ (Figs. S21e,f) and agree better with the experimental results (Figs. S21a,b). However, in the intermediate $n_e$ regime, QMC results obtained with $\varepsilon = 2.58\varepsilon_0$ (Figs. S21k,l) show stronger electron localization compared to the experimental results (Figs. S21c,d). QMC results obtained with $\varepsilon = 3.73\varepsilon_0$ (Figs. S21g,h) show a similar amount of local melting compared to the experimental results in the intermediate $n_e$ regime (Figs. S21c,d).

The simulation cell in our QMC simulations was chosen to mimic the experimental setup as closely as possible. We mapped the $140$ nm $\times$ $70$ nm field of view onto a $2 \times 1$ rectangular simulation box using an effective mass of $m_e^* = 0.54 m_e$ and an assumed dielectric constant. For example, with $\varepsilon = 3.73\varepsilon_0$, effective electron density $n_e^* = 4.7 \times 10^{11}$ cm$^{-2}$ for the LDD simulation translates to $r_s \approx 22.5$. This is realized by putting 32 electrons in a box of size $383.5a_0^* \times 191.75a_0^*$. Charged defects visible in the field of view are placed at their corresponding locations in the simulation cell.

|  | e | f | g | h | i |
| --- | --- | --- | --- | --- | --- |
| Fig 2 | $-0.876687 \pm 0.000008$ | $-1.11497 \pm 0.00001$ | $-1.236391 \pm 0.000005$ | $-1.613179 \pm 0.000006$ | N/A |
| Fig 3 | $-2.001260 \pm 0.000005$ | $-2.682208 \pm 0.000008$ | $-3.108614 \pm 0.000009$ | $-3.70128 \pm 0.00001$ | N/A |
| Fig 4 | $-3.80926 \pm 0.00005$ | $-4.36438 \pm 0.00003$ | $-7.2016 \pm 0.0001$ | $-11.2110 \pm 0.0003$ | N/A |
| Fig 5 | N/A | $-4.86207 \pm 0.00003$ | $-4.847 \pm 0.005$ | $-4.893 \pm 0.008$ | $-4.9383 \pm 0.008$ |

**Table I:** Energies in effective Hartree unit ($\text{Ha}^* = \hbar^2/m^* a_B^*$) for the simulations plotted in Figures 2, 3, 4, and 5 of the main text.

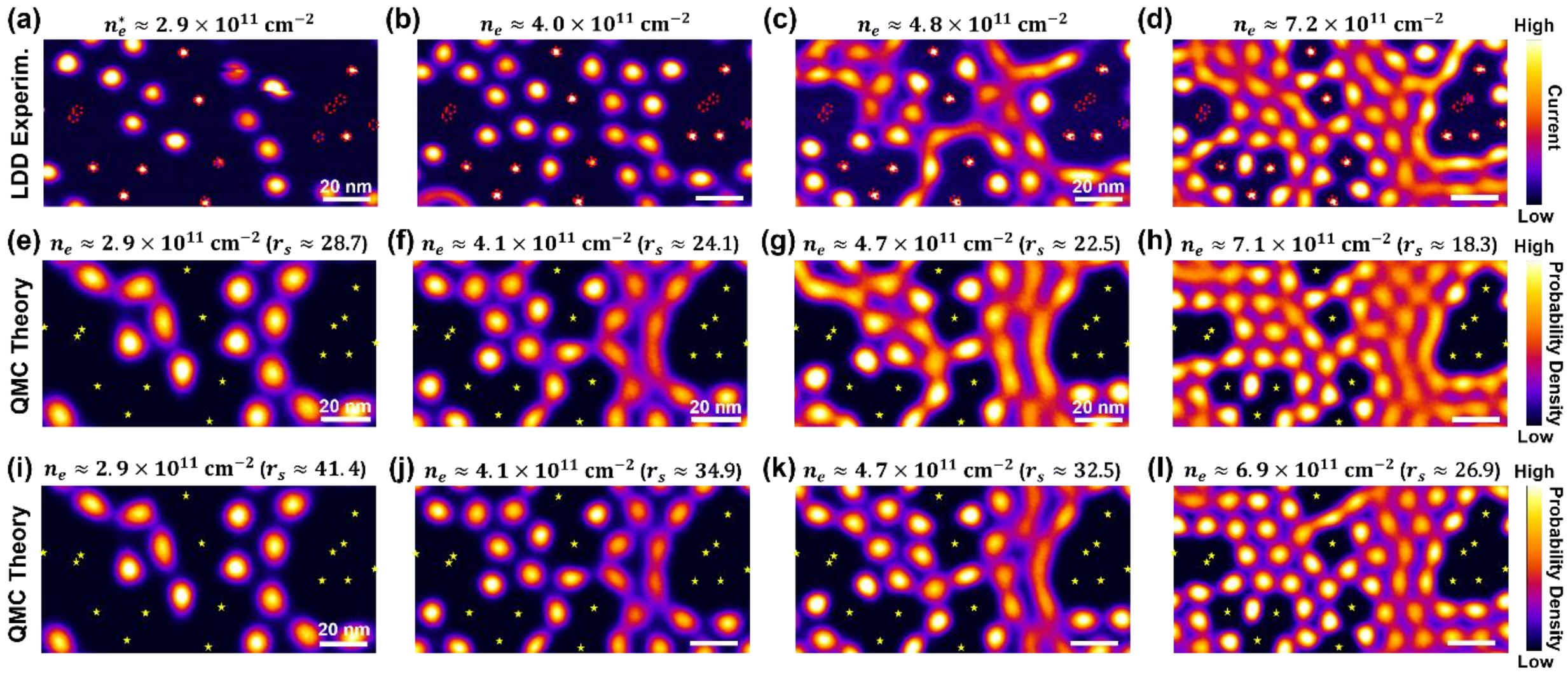

**Figure S21: Comparison between experimental and QMC simulated electron distributions with different $\varepsilon$. a-d,** Experimental in-gap tunnel current maps measured at (a) $V_G = 1.60$ V, (b) $V_G = 3.30$ V, (c) $V_G = 4.50$ V, (d) $V_G = 8.15$ V. Setpoint: $V_S = -1.00$ V. Estimated experimental $n_e$ values are shown for each image. **e-h,** Theoretical QMC simulations of electron probability distribution for (e) $n_e = 2.9 \times 10^{11}$ cm$^{-2}$, (f) $n_e = 4.1 \times 10^{11}$ cm$^{-2}$, (g) $n_e = 4.7 \times 10^{11}$ cm$^{-2}$, (h) $n_e = 7.1 \times 10^{11}$ cm$^{-2}$. Theoretical $r_s$ values are shown for each image. Dielectric constant used in QMC simulations for (e)-(h) is $\varepsilon = 3.73\varepsilon_0$. **i-l,** Theoretical QMC simulations of electron probability distribution for (i) $n_e = 2.9 \times 10^{11}$ cm$^{-2}$, (j) $n_e = 4.1 \times 10^{11}$ cm$^{-2}$, (k) $n_e = 4.7 \times 10^{11}$ cm$^{-2}$, (l) $n_e = 7.1 \times 10^{11}$ cm$^{-2}$. Theoretical $r_s$ values are shown for each image. Dielectric constant used in QMC simulations for (i)-(l) is $\varepsilon = 2.58\varepsilon_0$.

## S15. QMC simulation results for short-range disorder in the LDD regime

We performed QMC simulations in the LDD regime with short-range disorder to check that there are no significant effects. In the LDD regime the short-range disorder density is roughly equal to the long-range disorder density. Given the location of the isovalent defects, we model half as attractive and half as repulsive short-range disorder potentials. The resulting electron distributions with short-range disorder are nearly identical to the simulation results without short-range disorder (Fig. S22). The most noticeable effect is a slight reduction in local melting at $n_e \approx 3.3 \times 10^{11}\ \mathrm{cm}^{-2}$ as shown in Fig. S22f.

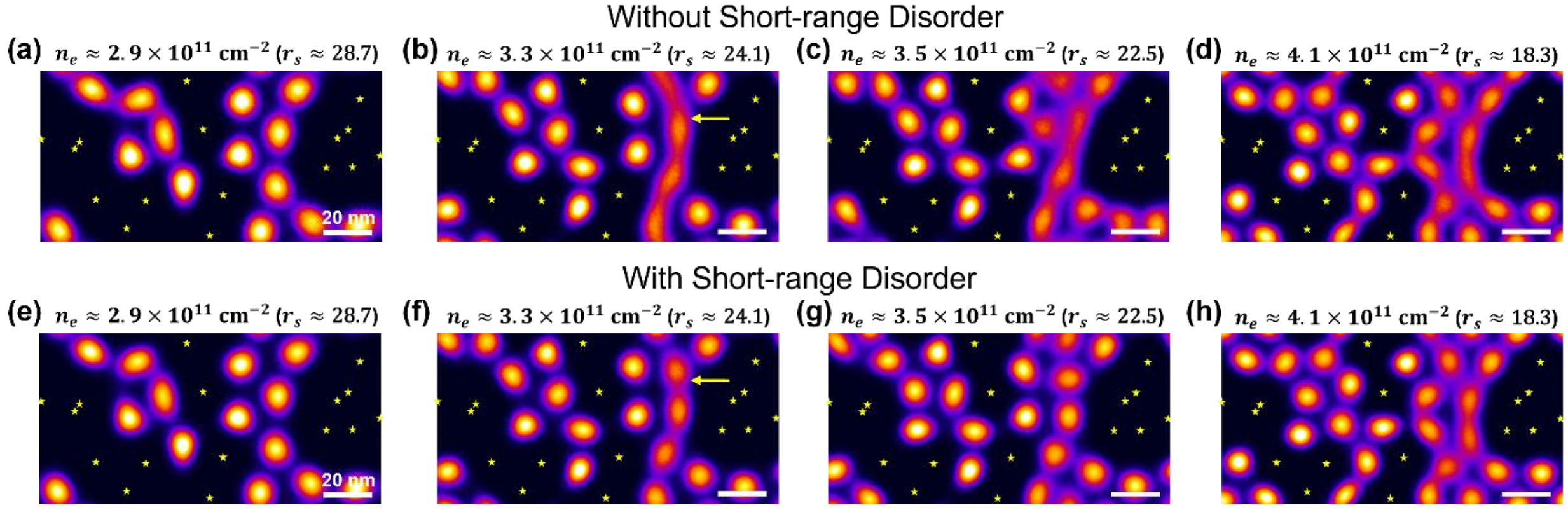

**Figure S22: Comparison between QMC simulations in the LDD regime with and without including short-range disorder. a-d,** QMC simulated electron probability distribution without including short-range disorder for (a) $n_e = 2.9 \times 10^{11}\ \mathrm{cm}^{-2}$, (b) $n_e = 3.3 \times 10^{11}\ \mathrm{cm}^{-2}$, (c) $n_e = 3.5 \times 10^{11}\ \mathrm{cm}^{-2}$, and (d) $n_e = 4.1 \times 10^{11}\ \mathrm{cm}^{-2}$. **e-h,** QMC simulated electron probability distribution with included short-range disorder for the same corresponding $n_e$ values as in (a)-(d). Yellow stars mark the locations of charged defects obtained from experimental positions and used in the simulations. Yellow arrows in (b) and (f) mark regions exhibiting local electron melting. Dielectric constant used in these QMC simulations is $\varepsilon = 3.73\varepsilon_0$.

**S16. Commensuration effect between electron WS and long-range disorder distribution**

To gain deeper insight into the local re-entrant melting/crystallization behavior observed for electron WSs in the LDD regime, we compared the structure factor of the electron density distribution with that of the fixed long-range disorder distribution. Figure S23a shows the charged defect positions for the measurement window shown in Figs. 2-4. To get the structure factor $s(k)$ of the disorder distribution, we first calculate the autocorrelation function of the disorder distribution shown in Fig. S23a. The resulting autocorrelation map is shown in Fig. S23b. We then perform a fast Fourier transform (FFT) of the autocorrelation function in Fig. S23b to get $s(k)$, which is shown in Fig. S23c (here we actually plot the squared structure factor $|s(k)|^2$ to enhance the $s(k)$ peak features). Long-range disorder peaks in $|s(k)|^2$ are circled by blue lines in Fig. S23c.

We then overlay the long-range disorder $|s(k)|^2$ peaks onto the electron density $|s(k)|^2$ map acquired through a similar process for in-gap tunnel current maps. Figures S24a-d show the same in-gap tunneling current maps shown in Figs. 2a-d. The corresponding electron density $|s(k)|^2$ maps with long-range disorder $|s(k)|^2$ peaks overlaid on top are shown in Figs. S24e-h. We observe robust locking between some of the electron and disorder $|s(k)|^2$ peaks (see yellow arrows in Figs. S24e,h) when the electron WS is in a more well-ordered state (Figs. S24a,d), and weaker locking when the electron WS has local melting (Figs. S24b,c). For the well-ordered electron WSs at two different $n_e$ values in Figs. S24a,d, the electron $|s(k)|^2$ peaks are seen to lock to different disorder $|s(k)|^2$ peaks. This indicates the experimentally observed local re-entrant melting/crystallization is related to a commensuration-incommensuration transition between electron and long-ranged disorder distributions as the electron WS transitions between different stable configurations while varying $n_e$.

Such defect locking between the electron WS and long-range disorder is also seen in our QMC simulations. Figures S25a-d show the same QMC simulated electron density distribution shown in Figs. 2e-h. Their corresponding $|s(k)|^2$ maps with long-range disorder $|s(k)|^2$ peaks overlaid on top are shown in Figs. S25e-h. Robust locking between the electron and disorder $|s(k)|^2$ peaks is observed at $n_e \approx 4.0 \times 10^{11}$ cm$^{-2}$ (see yellow arrows in Fig. S25h) where the electron WS is in a well-ordered state, while the locking disappears when the WS has local melting (Figs. S25b). Although the simulated WS phase shows no local melting at $n_e \approx 2.9 \times 10^{11}$ cm$^{-2}$ and $3.5 \times 10^{11}$ cm$^{-2}$ (Figs. S25a,c), there is no clear locking between the electron and disorder $|s(k)|^2$ peaks (Figs. S25e,g) as seen in our experiment. This discrepancy is likely due to the periodic boundary conditions applied in our QMC simulations, which have a larger effect at lower $n_e$. But the *orientations* of the electron $|s(k)|^2$ peaks are locked by the disorder $|s(k)|^2$ peaks as seen in Figs. S25e,g.

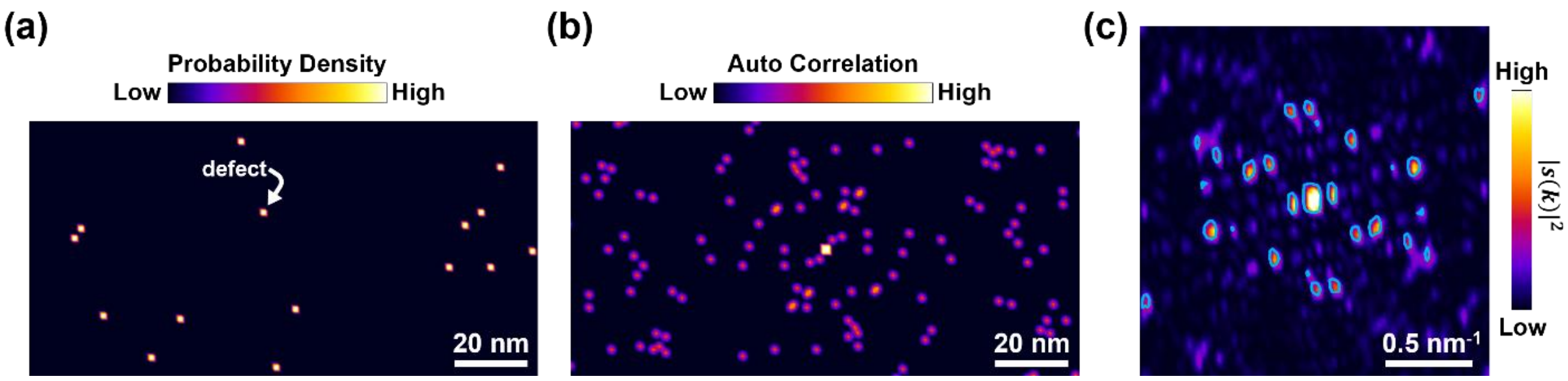

**Figure S23: Disorder structure factor for LDD region. a,** Extracted long-range disorder positions for the region shown in Fig. 1b. **b,** Autocorrelation of the disorder distribution map shown in (a). **c,** Squared structure factor of the disorder distribution map shown in (a), which is generated by generating squared FFT of the autocorrelation shown in (b). The blue circles mark the positions of $|s(k)|^2$ peaks.

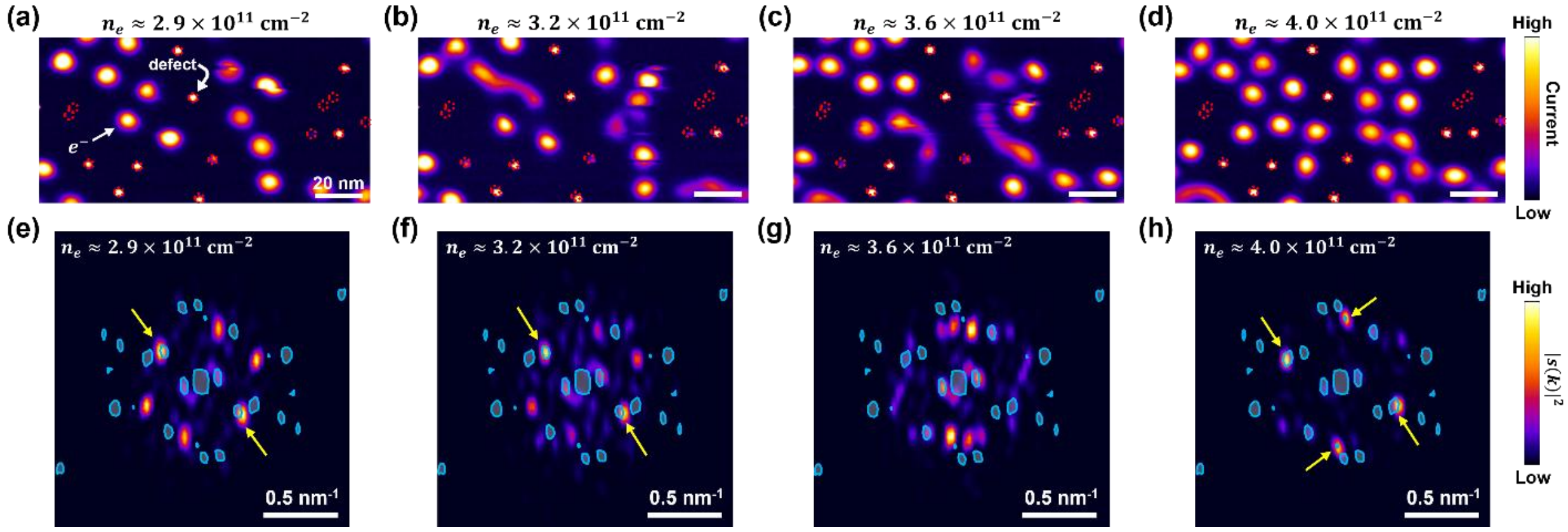

**Figure S24: Commensuration between WS electron distribution and disorder distribution in LDD experiment. a-d,** In-gap tunnel current maps from Figs. 2a-d. **e-h,** The corresponding $|s(k)|^2$ of the electron WS shown in (a)-(d). The blue circles represent the disorder $|s(k)|^2$ peaks for this region, which are extracted from Fig. S23c. The yellow arrows mark electron WS $|s(k)|^2$ peaks that overlap with disorder $|s(k)|^2$ peaks.

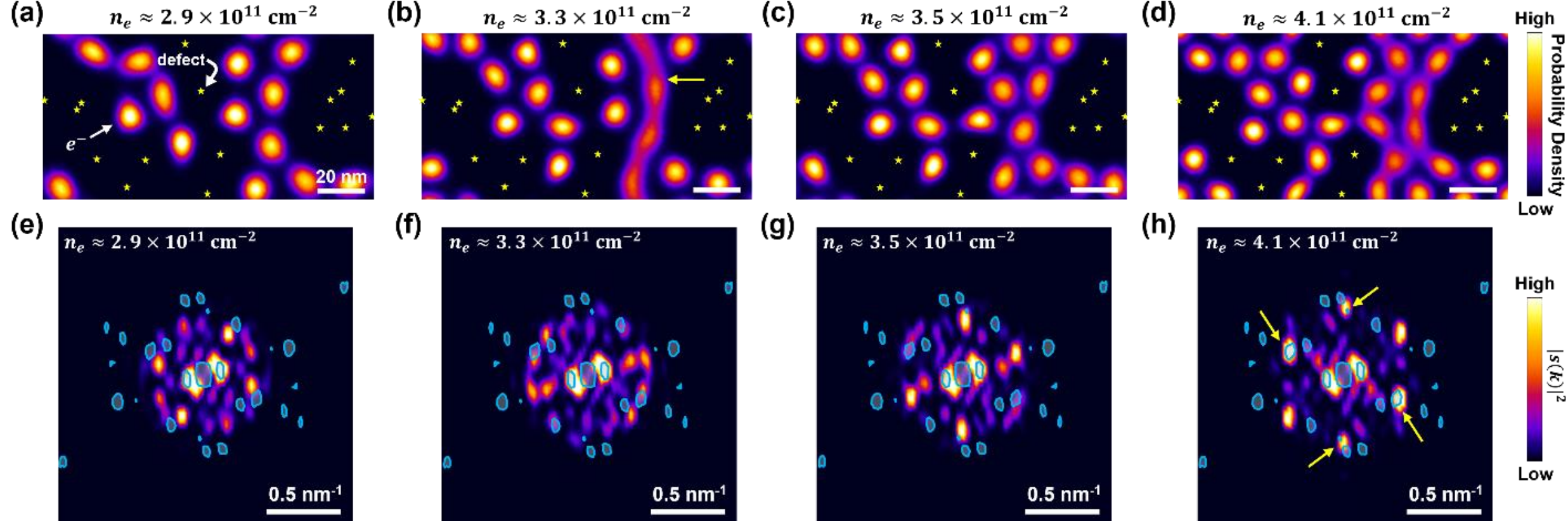

**Figure S25: Commensuration between WS electron distribution and disorder distribution in LDD simulation. a-d,** QMC simulated electron distribtuion maps from Figs. 2e-h. **e-h,** The corresponding $|s(k)|^2$ of the electron WS shown in (a)-(d). The blue circles represent the disorder $|s(k)|^2$ peaks for this region, which are extracted from Fig. S23c. The yellow arrows mark electron WS $|s(k)|^2$ peaks that overlap with disorder $|s(k)|^2$ peaks.

The local re-entrant melting/crystallization phenomena and commensuration effect between electron and disorder distributions mentioned above is highly reproducible. We observe it in the LDD regime every time when the STM tip and sample are free of contamination (more than 8 regions from 3 LDD BL-$MoSe_2$ devices). In Figs. S26 and S28 we show $n_e$-dependent electron Wigner solid images that show local re-entrant melting/crystallization behavior acquired from two additional LDD regions (region #1 and region #2). We performed an analysis similar to that shown in Fig. S24 to analyze commensuration between the WS electron distributions and disorder distributions in these two regions. The results for the additional LDD region #1 are shown in Fig. S27. The charged defect distribution in this region gives rise to disorder $|s(k)|^2$ peaks that are marked by blue circles/dots in Fig. S27. As $n_e$ increases the overlap between the disorder $|s(k)|^2$ peaks and the WS electron $|s(k)|^2$ peaks (marked by yellow arrows) changes. This coincides with local re-entrant melting/crystallization in real space images as $n_e$ increases (Fig. S26). The local re-entrant behavior is correlated with local commensuration-incommensuration transitions between the electron and disorder distributions.

LDD region #2 shows a smaller region with a clearly triangular charged defect distribution (Fig. S28). This gives rise to a triangular lattice $|s(k)|^2$ disorder peak distribution (blue circles in Fig. S29). Here the commensuration between electron and disorder distributions is evidenced by the strong overlap between the electron and disorder $|s(k)|^2$ peaks (marked by yellow arrows in Fig. S29). The 2D electrons exhibit a more solid phase when the overlap between electron and disorder $|s(k)|^2$ peaks is stronger (Figs. S28a,c,e and Fig. S29a,c,e). These results further support the interpretation that the local re-entrant phenomena for electron WSs in real space is related to a commensuration-incommensuration transition between electron and disorder distributions in reciprocal space.

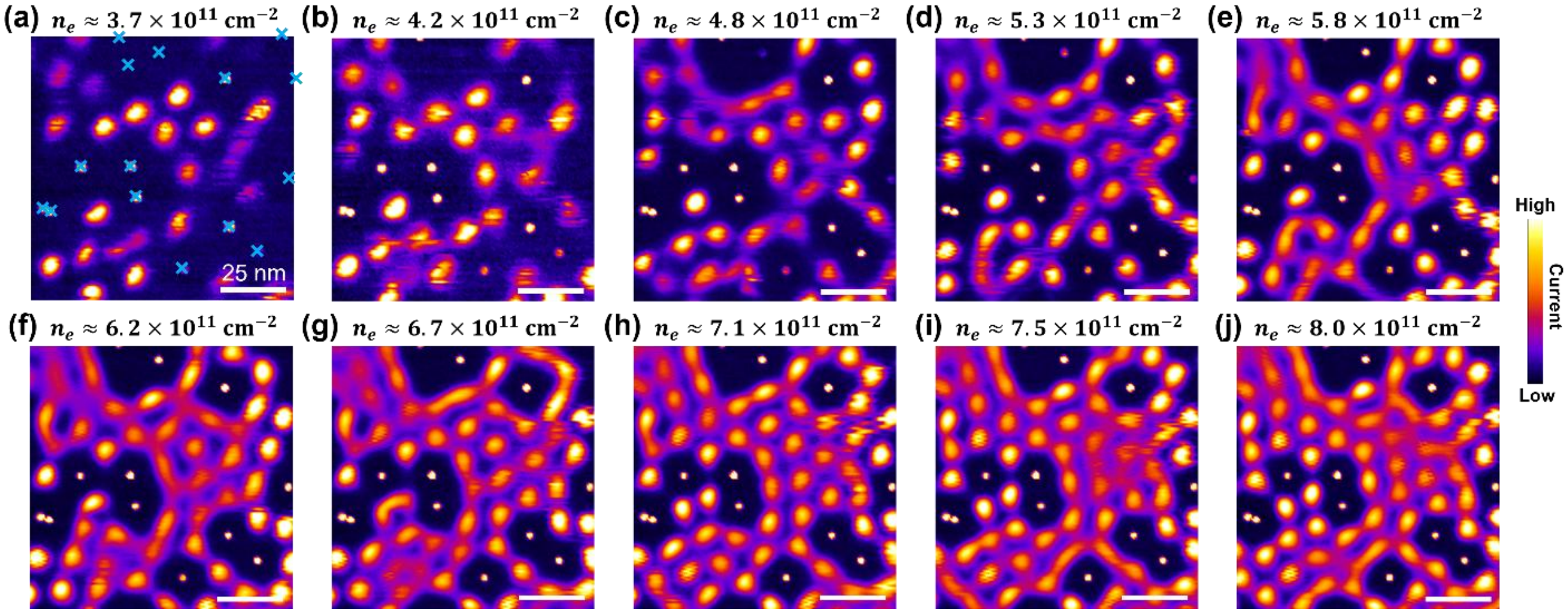

**Figure S26: Local re-entrant melting/crystallization behavior in additional LDD region #1.** Experimental in-gap tunnel current maps of an LDD BL-$MoSe_2$ device measured at (a) $V_G = 7.0$ V, (b) $V_G = 7.5$ V, (c) $V_G = 8.0$ V, (d) $V_G = 8.5$ V, (e) $V_G = 9.0$ V, (f) $V_G = 9.5$ V, (g) $V_G = 10.0$ V, (h) $V_G = 10.5$ V, (i) $V_G = 11.0$ V, and (j) $V_G = 11.5$ V. Setpoint: $V_S = -1.00$ V. Experimental estimated $n_e$ values are shown. Blue crosses mark the locations of charged defects.

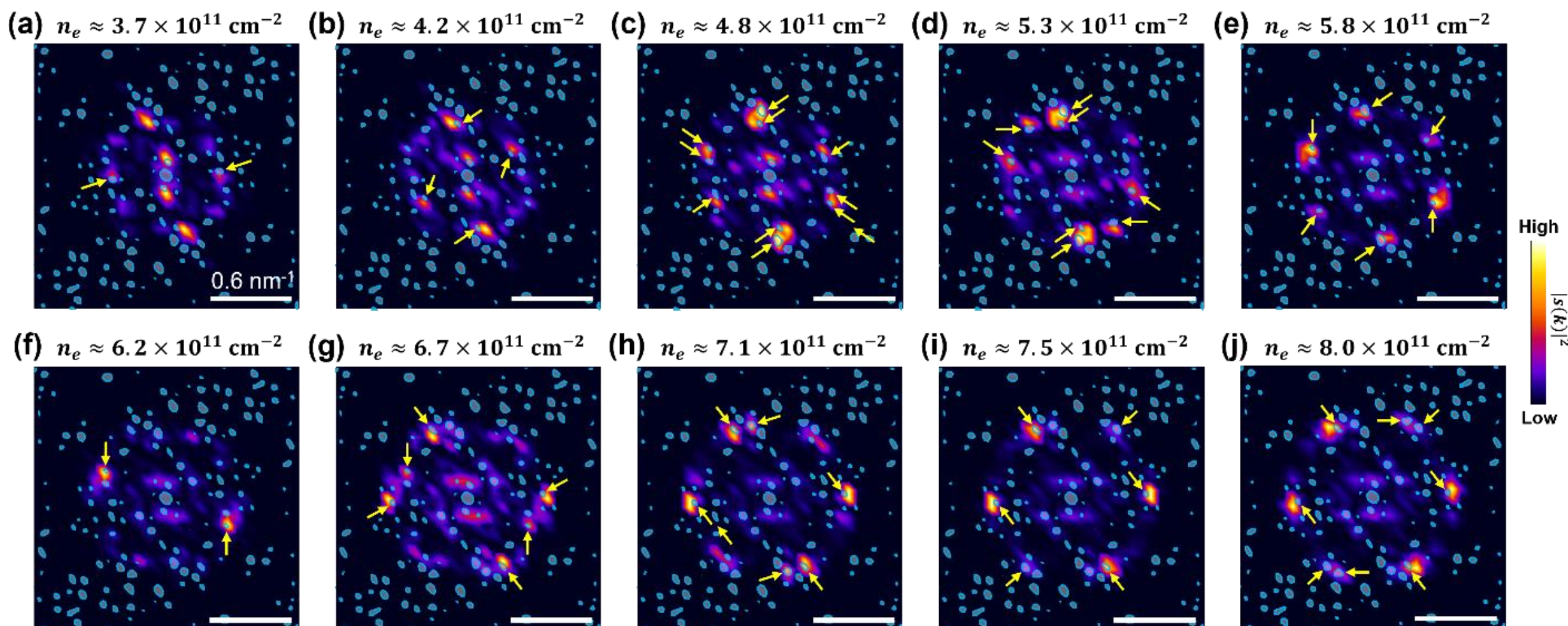

**Figure S27: Commensuration between WS electron distribution and disorder distribution in additional LDD region #1. a-j,** $|s(k)|^2$ for the electron WS images shown in Figs. S26a-j. The blue circles represent the disorder $|s(k)|^2$ peaks for the charged defect distribution in this region. The yellow arrows mark where the disorder $|s(k)|^2$ peaks overlap with the electron WS $|s(k)|^2$ peaks.

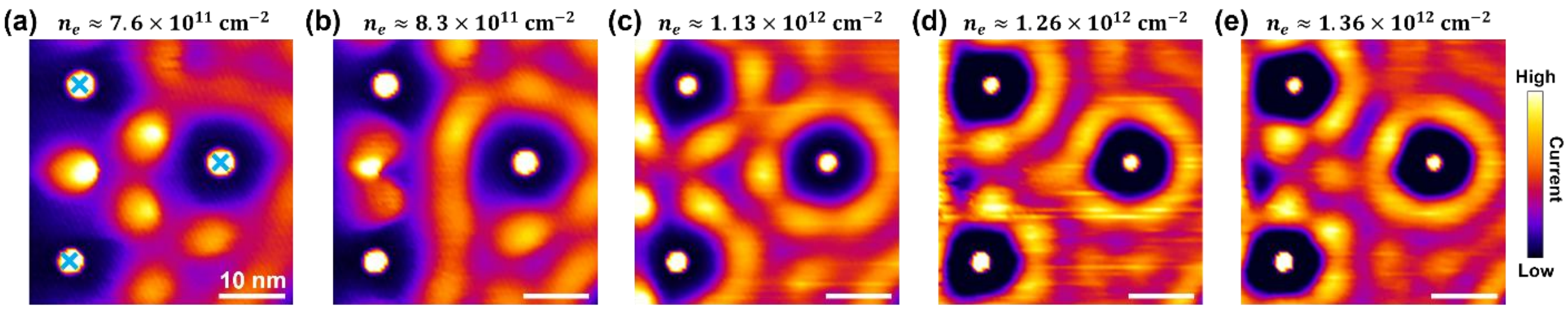

**Figure S28: Local re-entrant melting/crystallization behavior in additional LDD region #2.** Experimental in-gap tunnel current maps of an LDD BL-$MoSe_2$ device measured at (a) $V_G = 0.0$ V, (b) $V_G = 1.0$ V, (c) $V_G = 5.2$ V, (d) $V_G = 7.0$ V, and (e) $V_G = 8.4$ V. Setpoint: $V_S = -1.00$ V. Experimental estimated $n_e$ values are shown (The doping of BL-$MoSe_2$ was offset here by an STM tip-pulsing technique, thus $n_e$ is non-zero at $V_G = 0$ V). Blue crosses mark the locations of charged defects.

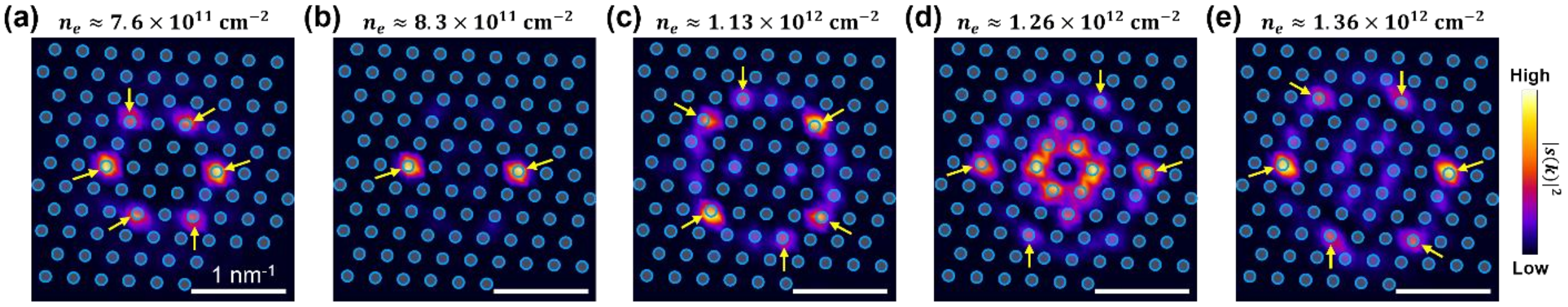

**Figure S29: Commensuration between WS electron distribution and disorder distribution in additional LDD region #2. a-e,** $|s(k)|^2$ for the electron WS images shown in Figs. S28a-e. The blue circles represent the disorder $|s(k)|^2$ peaks for the charged defect distribution in this

region. The yellow arrows mark where the disorder $|s(k)|^2$ peaks overlap with the electron WS $|s(k)|^2$ peaks.

**S17. Effect of increasing electron screening on HDD local electron density**

In Figs. 5f-i of the main text we showed that increasing $n_{SR}$ enhances electron localization and the formation of WSs in the HDD regime. To show that this effect arises from electron-electron interactions we used QMC to simulate the HDD device with a full complement of isovalent defects, but for increasingly strong electronic screening. The idea here is to see if the HDD WS behavior reduces as we reduce electron-electron interactions between charge carriers through increased screening. Screening was added to our HDD 2DEG simulations by incorporating an additional dual gate configuration to the simulation, thus allowing us to gradually "turn off" electron-electron interactions between charge carriers by decreasing the distance between the 2DEG and the added gates (interactions between charge carriers and defects was left unchanged). Figure S30a shows a QMC simulation of the HDD device shown in Fig. 5a with the full complement of isovalent defects, an electron density of $n_e \approx 7.8 \times 10^{11}$ cm$^{-2}$, and *no* added screening of charge carriers (i.e., the added gates are very far away, $d \approx 30.4$ nm for this simulation). The amorphous WS characteristics of the HDD regime are clearly seen here in the absence of strong screening. As screening is added by bringing the gates closer to the 2DEG, however, electron localization in the HDD WS phase is seen to diminish (Fig. S30b,c), and for very strong screening the HDD WS phase completely disappears (Figs. S30d). The presence of robust electron-electron interactions between charge carriers is thus seen as an essential component leading to the observed behavior of the WS phase in the HDD regime.

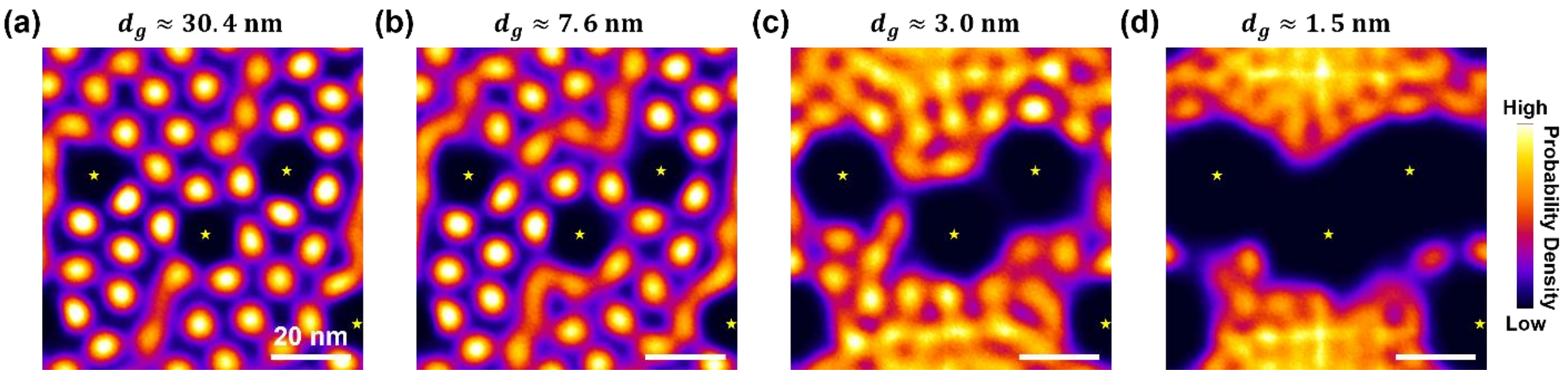

**Figure S30: Effect of screening.** Local electron density is shown in the HDD regime for QMC simulations having increasing amounts of screening for charge carriers ($n_e \approx 7.8 \times 10^{11}$ cm$^{-2}$, $r_s \approx 21.0$ is fixed). Screening here arises from the addition of dual gates in the simulation placed at a distance $d_g$ from the HDD 2DEG layer (the reduction of $d_g$ leads to an increase in charge carrier screening). Dielectric constant used in simulations is $\varepsilon = 3.10\varepsilon_0$, and $n_{SR} = n_0 \approx 7.63 \times 10^{12}$ cm$^{-2}$.

## S18. Quenching of Friedel oscillations by short-range disorder in QMC simulations of HDD regime

The quenching of Friedel oscillations around charged defects as $n_{SR}$ is increased from the LDD to the HDD regime is also seen in our QMC simulations. Figures S31a-c show the QMC simulated electron density distribution around a single charged defect at $n_e \approx 2.54 \times 10^{12}$ cm$^{-2}$ ($r_s \approx 14.0$) (similar to $n_e$ of Fig. 5e) with increasing $n_{SR}$. Ring-like Friedel oscillations can be clearly observed around the defect at $n_{SR} = 0$ (Fig. S31a) but become quenched at $n_{SR} \approx 5.43 \times 10^{12}$ cm$^{-2}$ (Fig. S31b) and $n_{SR} \approx 1.086 \times 10^{13}$ cm$^{-2}$ (Fig. S31c).

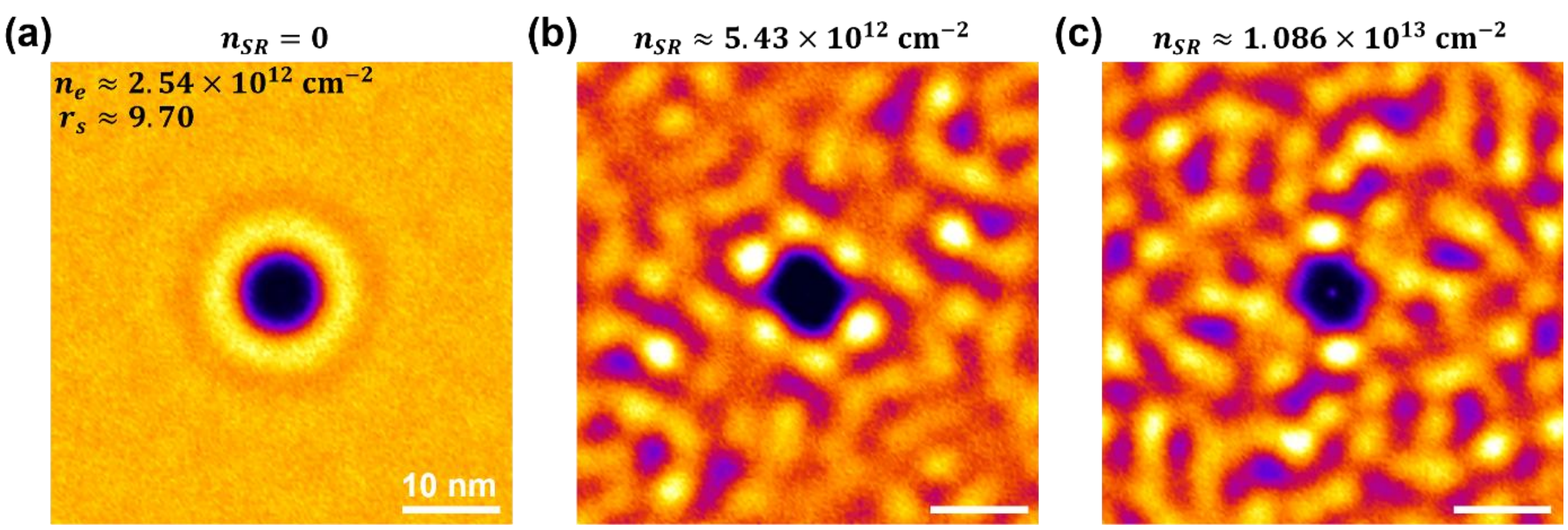

**Figure S31: Quenching of Friedel oscillations by increasing short-range disorder density. a-c,** QMC simulated electron density distribution around a single charged defect at $n_e \approx 2.54 \times 10^{12}$ cm$^{-2}$ ($r_s \approx 9.72$) with (a) $n_{SR} = 0$, (b) $n_{SR} \approx 5.43 \times 10^{12}$ cm$^{-2}$, and (c) $n_{SR} \approx 1.086 \times 10^{13}$ cm$^{-2}$. Dielectric constant used in QMC simulations for (a)-(c) is $\varepsilon = 3.73\varepsilon_0$.